\documentclass[11pt]{article}
\usepackage{hyperref}
\usepackage{cite}
\usepackage{graphicx}
\usepackage{subfigure}
\usepackage[top=3cm, bottom=3cm, left=1.5cm, right=1.5cm]{geometry}
\usepackage{amsmath,amsfonts,amssymb,mathtools}
\usepackage{graphicx}
\usepackage{float}
\usepackage{breqn}
\usepackage{multicol}
\usepackage{multirow}
\usepackage{booktabs}
\usepackage{xcolor}
\usepackage{array,makecell, cellspace}
\date{}

\begin{document}
	
	\setcounter{page}{1}
	\pagestyle{plain}

	\begin{center}
		\Large{\bf  Non-minimal Unimodular Inflation}\\
		\small \vspace{1cm} {\bf Manda
			Malekpour\footnote{m.malekpour@stu.umz.ac.ir}},\,\,{\bf Kourosh
			Nozari\footnote{knozari@umz.ac.ir(Corresponding Author)}},\,\, {\bf Fateme Rajabi\footnote{fa.rajabi@stu.umz.ac.ir}}\quad and \quad {\bf Narges
			Rashidi\footnote{n.rashidi@umz.ac.ir}},
		\\
		\vspace{0.25cm}
		Department of Theoretical Physics, Faculty of Sciences,\\
		University of Mazandaran,\\
		P. O. Box 47416-95447, Babolsar, IRAN
	\end{center}

\begin{abstract}
	
We study an extension of the unimodular cosmological inflation in the context of the non-minimal coupling of a generic scalar field with gravitational sector. We consider the non-minimal coupling of the scalar field and gravity as the only source of energy-momentum tensor in this setup. Without introducing new particles other than those already existing in electroweak theory, the generic scalar field's non-minimal coupling is responsible for the generation of the seeds of perturbations for structure formation and the observed Cosmic Microwave Background anisotropies in this scenario. We calculate inflation parameters in both the Jordan and Einstein frames, and then we study primordial spectral indices for slow-roll parameters in Einstein's frame at the first and second orders. The numerical results of this model are consistent with the Planck2018 and BICEP/Keck joint data sets in some subspaces of the model parameters space. By considering a sufficient amount of inflation, we estimate the strength of the non-minimal coupling parameter, $\xi$, to find appropriate new constraints on the values of this parameter. By comparing the numerical values of the inflation observables in two frames and also with observation, we comment on the issue of frames in this framework.
\end{abstract}

\newpage
 \section{Introduction}

 Nowadays, one of the important issues in modern cosmology is the  \textit{\textquotedblleft cosmological inflation\textquotedblright}, which was originally proposed by Alan Guth in 1981 \cite{guth1981inflationary}, to explain the initial conditions of the hot big bang model. Inflation takes place after the Big Bang during a time interval that is believed to be approximately given as $ t \sim  10^{-36}- 10^{-32}$ sec \cite{Sebastiani:2015kfa}. A de Sitter expansion in the inflationary period can address the flatness, horizon, and monopole problems of the standard hot Bing Bang cosmology \cite{Rio02,baumann2018tasi} and can also provide seeds for the temperature anisotropy of the Cosmic Microwave Background (CMB) in the present universe. The inflationary model explains how the initial seeds of the large-scale structure are formed by considering the disturbances that occur during the inflationary period due to quantum fluctuations \cite{starobinsky1979relict,wang2014inflation,senatore2016lectures, Martin:2013tda}.

The simplest model of inflation is described by a single canonical scalar field, the so-called inflaton, that slowly rolls down from the peak of a self-interacting potential to the minimum point of the potential in the context of the slow-roll approximation. Various inflationary scenarios have been proposed that could explain the accelerated expansion of the early universe. The observational constraints of Planck data have restricted or ruled out a wide range of single-field models \cite{martin2016have}. It is believed that inflation is driven by a scalar field that is essentially non-minimally coupled to other fields, such as gravity. Among the widely used inflationary models is the Non-Minimal Coupling (NMC) of the scalar field with gravity, which is essential in many situations of physical and cosmological interest. NMC arises in the presence of the scalar field quantum corrections. It is also necessary to renormalize the scalar field theory in curved space. It seems that in most inflationary theories, the non-vanishing value of the NMC parameter is inevitable \cite{M. Demia,faraoni1998conformal,faraoni2000inflation,Nozari:2007eq,park2008inflation,nozari2010non,hertzberg2010inflation,qiu2012reconstruction,shaposhnikov2013cosmology,chiba2015consistency,Myrzakul:2015gya,bostan2019quartic}. NMC terms in inflation may lead to corrections on the power spectrum of primordial perturbations
\cite{salopek1989designing,fakir1990improvement,kaiser1995primordial,hochberg1995energy} and Non-Gaussianities \cite{qiu2011non}, as well as a tiny tensor-to-scalar ratio \cite{komatsu1998constraints,komatsu1999complete,hwang1999cobe}, which puts constraints on the parameters of the model. Furthermore, Fakir and Unruh demonstrated that the fine-tuning problem of the self-coupling in their setup is relaxed when the amplitude of the scalar metric perturbations takes large negative values. As a result of their nontrivial renormalization group flows, quantum field theory in curved spacetime naturally exhibits NMC \cite{tsujikawa2004density}. In NMC theory, we are generally free to choose any form of action, especially coupling and its potential conditions. To avoid the complexity of NMC, the theory can be transformed from the Jordan frame to the Einstein frame since they are often used as a mathematical tool to map the equations of motion into at least mathematically equivalent sets of equations that are more easily solved and computationally more convenient to study.

The fundamental candidate for the inflaton is the Standard Model's Higgs boson. The scalar potential of the Higgs boson in the Standard Model of particle physics behaves asymptotically like the self-interacting quartic potential $V (\phi) = {\lambda_{0}}{\phi}^4/4$ in renormalizable gauge field theories \cite{Pich:2007vu,olive2016review}. But this potential suffers from some critical problems in the standard inflationary era. The Higgs boson minimally coupled to gravity is inconvenient by the observational data due to its large tensor-to-scalar ratio \cite{linde1983chaotic,akrami2020planck}. The non-minimally coupled Higgs field to gravity solves this problem, and the tensor-to-scalar ratio $r$ could be reduced to be consistent with the observational data, and the Higgs effective self-coupling $\lambda_{0}$ could be of the order of $1$ \cite{germani2010new,germani2014self}. On the other hand, the issue of unitary can be addressed by inclusion of the non-minimal derivative coupling. Hence, the non-minimal derivative coupling inflation model provides the Higgs model without introducing a new degree of freedom. In the case of the Standard Model of the Higgs field, several authors investigated and discussed the effects of a NMC on the field inflation \cite{bezrukov2008non,Bezrukov:2009db,bezrukov2013higgs,saltas2016higgs,rubio2019higgs,granda2020higgs}. Also, in 2008, Bezrukov and Shaposhnikov \cite{Bezrukov:2007ep} investigated the Standard Model with a NMC term between the Higgs field and the Ricci scalar (in the form of $\xi |\phi|^2 R$) that could cause inflation, where the $ \xi $ parameter is the dimensionless coupling constant giving rise to the simplest extension of the scalar field Lagrangian. Notice that one has varieties of predictions of the Higgs inflation depending on the formulation of the background gravitational theory \cite{shaposhnikov2015higgs}. The inflaton's interaction with Standard Model fields is believed to be an essential factor in determining the success of inflation. In particular, it is still unclear how the inflaton's coupling with Standard Model fields affects the universe's evolution during and after inflation.
 
Historically, in 1919, Einstein proposed a theory to solve the cosmological constant problem, which is known as \textit{unimodular gravity} \cite{einstein1952principle}.
Several years later, Anderson and Finkelstein \cite{anderson1971cosmological} proposed the concept of unimodular gravity in 1970. Many aspects of this theory have been studied in \cite{gao2014cosmological,cho2015unimodular,Nojiri:2015sfd,Rajabi:2017alf,barvinsky2017darkness,barvinsky2019inflation,barvinsky2019dynamics,Herrero-Valea:2020xaq,Rajabi:2021bdd,Rajabi:2022qrs,leon2022inflation}. In this scenario, the cosmological constant appears in the theory as an integration constant \cite{weinberg1989cosmological,unruh1989unimodular,liddle1993end,sahni2002cosmological}. In other words, the cosmological constant arises as an arbitrary constant of motion rather than a fixed fundamental parameter in the action. Unimodular gravity does not require the cosmological constant term. Thus, this has the potential to solve the fine-tuning problem of the cosmological constant \cite{jain2012testing,jain2012cosmological}. The main idea of unimodular gravity is that the determinant of the metric $\sqrt{-g}$ is constrained to a constant number or a function of spatial coordinates \cite{nozari2017cosmological}. However, as Weinberg explained in \cite{weinberg1989cosmological}, the problem is not solved completely, and this is an open issue in this regard.

The motivation of this study is to see whether a generic, mathematical scalar function has the capability to derive cosmological inflation? For this purpose, we have not attributed any kinetic and potential terms to the generic scalar function; instead, we have assumed it has a NMC with the background curvature via a coupling that has been borrowed from non-minimal Higgs inflation. Then, since there is no trace of the kinetic and potential terms in the original action, we are going to give this role to the unimodularity of the model. That is, we think that a unimodular framework that naturally generates a cosmological constant may compensate for the lack of potential in the Jordan frame (since the Einstein frame has an effective potential for this pure mathematical field). In choosing the NMC, we consider the coupling in the Higgs non-minimal inflation as $ f(\phi) = \frac{M_{P}^{2}}{2} \left( 1+\frac{\xi \phi^2}{M_{P}^{2}}\right) $. Since in the Jordan frame we discard the existence of a potential of the generic field, we use the Hubble Slow-Roll conditions. In the Einstein frame, we have effective potential as $U_{eff}(\varphi)$, and therefore Potential Slow-Roll parameters are applicable as usual. About why we consider unimodular framework, in addition to the reason stated above, it is well known that at the classical level, unimodular gravity produces the same physics as general relativity with a cosmological constant. However, the difference between them is that the cosmological constant of unimodular gravity is a constant of integration, while in general relativity it is a coupling constant. In other words, the vacuum energy does not couple to gravitation in a unimodular framework. We have shown that a generic scalar field with NMC to gravity in a unimodular framework is capable of achieving feasible cosmological inflation. This is the essence of this work in the Jordan frame. We note that since the value of the cosmological constant is unspecified and unrelated to any coupling constant, problems associated with the cosmological constant have been reconsidered in unimodular gravity in the literature. 

As we will show, this pure NMC has the potential to initiate a phase of slow-roll inflation, which could have significant implications for the evolution of the universe.
We calculate the inflation observables in the slow-roll approximation in both Jordan and Einstein frames in detail. We compare our numerical results with Planck2018 \cite{Planck:2018vyg,akrami2020planck} and BICEP/Keck \cite{ade2021improved,Paoletti:2022anb} joint data to check the consistency of the constructed model. In this manner, we find some new constraints on the NMC parameter, $\xi$. By comparing the numerical results in two frames in confrontation with observation, we comment on the issue of frames in this context.

The paper is structured as follows: In Sec. \ref{sec2}, we present the equations of motion in the Jordan frame with the unimodular constraint in a spatially flat Friedmann-Robertson-Walker (FRW) space-time. Next, in Sec. \ref{sec3}, we introduce a generic scalar field without kinetic and potential terms, but with NMC with gravitational sector as the inflaton and derive the equations of motion in the Jordan frame. In Sec. \ref{sec4}, we calculate the Hubble slow-roll parameters such as the spectral index $n_{s}$ and the tensor-to-scalar ratio $r$. Then we consider the ansatz $\lambda= A \phi^n $ for the Lagrange multiplier, $\lambda$, as a power law in the inflaton field to set constraints on $\xi$. In Sec. \ref{sec5}, we solve the Einstein frame field equations using the potential-slow-roll formalism. Sec. \ref{sec6} is devoted to studying the numerical results in both frames and comparing them with observational data. The paper terminates with a summary and conclusions in Sec. \ref{sec7}. Except for otherwise stated explicitly, we consider the units as $ c = \hbar = M_{P} =1$ and the metric signature as (-, +, +, +).

\section{Generalized Unimodular Gravity}\label{sec2}

The original motivation of unimodular gravity was to solve the cosmological constant problem by disentangling the cosmological constant from the gravitational equations of motion, or in other words, to make its contribution to be zero by requiring that the determinant of the spacetime metric is not dynamical, but is restricted to
\begin{equation}\label{eq1}
	\sqrt{-g}=\epsilon_{0}\,,
\end{equation}
where $\epsilon_{0}$ is a constant parameter. We consider a modest extension of the standard general relativity with the following action	
\begin{equation}\label{eq2}
	S_{J}= \int d^{4}x \left(\sqrt{-g} f\left(\phi\right)R- \frac{2\lambda\left( \sqrt{-g}-1\right)}{\left( 2\kappa^{2}\right)} \right)\,,		
\end{equation}
where $ f\left(\phi\right) $ is an analytic function of the scalar field $\phi$, $ R= g^{\mu\nu} R_{\mu\nu} $ is the Ricci scalar, and $ \lambda$ is a Lagrange multiplier (not to be confused with $\lambda_{0}$ introduced in the previous section for the Higgs effective self-coupling). In this unimodular theory, the inflaton field is non-minimally coupled to gravity. In this work, we set $ \kappa^{2} \equiv 8 \pi G = M_{P}^{-2} $. Variation of the action \eqref{eq2} with respect to the metric $ g_{\mu\nu} $ leads to the field equations as
\begin{equation}\label{eq3}
		 -\frac{1}{2} g_{\mu\nu} f\left( \phi\right) R + f\left(\phi\right)  R_{\mu\nu} +
		 \left( \Box g_{\mu\nu}- \nabla_{\mu} \nabla_{\nu} \right) f\left( \phi\right) + \lambda  g_{\mu\nu}=0 \,,
\end{equation}
where $ \Box  \equiv  g^{\mu\nu} \nabla_{\mu} \nabla_{\nu} $ is d'Alembert's operator. When the action is varied with respect to $ \lambda$, the unimodularity condition \eqref{eq1} is established. The trace of equation \eqref{eq3} is given by
\begin{equation}\label{eq4}
	-2 f\left( \phi\right) R + R f\left( \phi\right)+ 3 \Box f(\phi) + 4 \lambda =0 \,.
\end{equation}
We consider a spatially flat FRW metric with a line element as
\begin{equation}\label{eq5}
	ds^2 = -dt^{2} + a^{2}\left( t\right) \sum_{i=1}^{3} \left( dx^{i}\right)^{2} \,,
\end{equation}
where $a\left( t\right)$ is the scale factor. This metric does not satisfy the unimodular constraint \eqref{eq1}. In this regard, we redefine the cosmic time coordinate as \cite{Nojiri:2015sfd}
\begin{equation}\label{eq6}
	d\tau = a^{3}\left(t\right) dt\,.
\end{equation}
Therefore with this new definition, the metric \eqref{eq5} can be rewritten as
\begin{equation}\label{eq7}
	ds^{2} = -a^{-6} \left( \tau\right)  d\tau^{2} + a^{2}\left( \tau\right)  \sum_{i=1}^{3} \left( dx^{i}\right)^{2} \,.
\end{equation}
Now, with the unimodular metric of equation \eqref{eq7}, we calculate the non-vanishing components of the Ricci tensor and also the Ricci scalar as
 \begin{equation}\label{eq8}
 	\begin{split}
 	 R_{\tau\tau}= -3\dot{\mathcal{H}} - 12 \mathcal{H}^2\,,
 	 \quad R_{ij}= a^8\left( \dot{\mathcal{H}} + 6\mathcal{H}^2 \right)  \delta_{ij}\,,
 	 \quad R= a^6 \left( 6 \dot{\mathcal{H}} +30 \mathcal{H}^2\right) \,,
 \end{split}
 \end{equation}
where $\mathcal{H} $ is the modified Hubble parameter as $ \mathcal{H}= \frac{1}{a}\frac{da}{d\tau} $ and a “dot” means differentiation with respect to the time parameter $\tau$. By using the equations \eqref{eq7} and \eqref{eq8}, the $\tau\tau$ and $ i j $ components of the field equations are given as follows
\begin{equation}\label{eq9}
	3 \mathcal{H}^2 f\left( \phi\right)  + 3\mathcal{H} \dot{f}\left( \phi\right)  - \lambda a^{-6}=0\,,
\end{equation}
\begin{equation}\label{eq10}
	-\left( 2\dot{\mathcal{H}}+9\mathcal{H}^{2} \right) f\left(\phi\right)-5 \mathcal{H} \dot{f} \left(\phi\right) -\ddot{f}\left( \phi\right)+\lambda a^{-6}=0\,.
\end{equation}
Now, by combining equations \eqref{eq9} and \eqref{eq10}, we find
\begin{equation}\label{eq11}
 	\ddot{f}\left(\phi\right)+2\mathcal{H} \dot{f}\left( \phi\right)+ \left(6 \mathcal{H}^{2}  + 2\dot{\mathcal{H}}\right)  f\left(\phi\right)=0\,,
\end{equation}
which essentially determines the time evolution of the NMC, $f\left( \phi\right) $.

\section{A Generic Scalar Field as Inflation}\label{sec3}

We consider the NMC of a generic, mathematical scalar function with a gravitational sector in the same fashion that usually happens in some non-minimal Higgs inflation contexts. In other words, we considered the NMC between the scalar field and the gravitational sector (the Ricci scalar) as
\begin{equation}\label{eq12}
	f(\phi) = \frac{1}{2} M_{P}^{2} \left( 1+  \frac{\xi \phi^2}{M_{P}^{2}}\right)\,
\end{equation}
which has been used in some non-minimal Higgs inflation contexts. On the other hand, the NMC term in the action effectively plays the role of a potential term for the generic scalar field in this setup. The main point here is that we conjecture that this coupling to gravity may be enough to realize cosmological inflation in the Jordan frame in this unimodular framework. In fact, we consider that the term $ \xi\phi^{2} R$ can play the role of an effective potential in the action, and since we are working in the slow-roll playground, the kinetic term is not so important in the realization of the cosmic inflation. But, since this assumption seemingly has some shortcomings in the foundation, we embedded the scenario in a unimodular framework with the hope of having parameter space wide enough to construct a feasible inflation model. In fact, we tried to embed a very simple non-minimal generic scalar field in the unimodular gravitational theory to see the possible outcomes in the realm of cosmological inflation. In this manner, in the Jordan frame, we have discarded to consider potential and kinetic terms for the generic scalar function, but we suppose it is coupled non-minimally to gravity within a unimodular framework with the hope of seeing the term $ \xi\phi^{2} R $ playing the role of a potential term and the unimodularity of the theory to add some more ingredient to the model in order to match its parameter space with observation. For these reasons, in the Jordan frame, in the lack of a potential, we use the Hubble Slow-Roll Parameters, whereas in the Einstein frame since we have an effective potential, we use the Potential Slow-Roll Parameters.
Thus, the action \eqref{eq2} can be rewritten as
\begin{equation}\label{eq13}
	S_{J}= \int d^{4}x  \biggl\{\sqrt{-g} \bigg[ \frac{1}{2} M_{P}^2 \left(1+  \frac{\xi\phi^2}{M_{P}^{2}}\right)\bigg] R- \frac{2\lambda\left( \sqrt{-g}-1\right)}{\left( 2\kappa^{2}\right) }\biggr\}\,,		
\end{equation}
where as usual $M_{P}$ is the Planck mass, $\phi$ is the scalar field, and $\xi$ is the dimensionless NMC parameter to be constraint later. To continue, by considering the action \eqref{eq13}, we calculate the field equations as follows
\begin{equation}\label{eq14}
	\frac{3}{2} M_{P}^{2} \mathcal{H}^{2} \left( 1+ \frac{\xi \phi^{2}}{M_{P}^{2}}\right)+ 3\xi \mathcal{H} \phi \dot{\phi} - \lambda a^{-6} =0\,,
\end{equation}
and
\begin{equation}\label{eq15}
	  -\frac{1}{2} M_{P}^{2} \left( 2\dot{\mathcal{H}} +9 \mathcal{H}^{2} \right) \left( 1+\frac{\xi \phi^{2}}{M_{P}^{2}}\right) - \xi \dot{\phi}^{2}
	  -\xi \phi \ddot{\phi} -5\mathcal{H} \xi \phi \dot{\phi} + \lambda a^{-6} =0\,.
\end{equation}
Then, by combining the equations \eqref{eq14} and \eqref{eq15} we obtain
\begin{equation}\label{eq16}
	 \frac{1}{2} M_{P}^{2} \left( -2\dot{\mathcal{H}} - 6 \mathcal{H}^{2} \right)   \left( 1+\frac{\xi \phi^{2}}{M_{P}^{2}}\right)
	 - 2 \xi\mathcal{H} \phi \dot{\phi}-\xi \dot{\phi}^{2} - \xi \phi \ddot{\phi}=0\,.	
\end{equation}
Now, we consider a power-law scale factor as
\begin{equation}\label{eq17}
	a\left(t\right)= \left( \frac{t}{t_{0}}\right)^{b}   \Longrightarrow H = \frac{b}{t} \,,
\end{equation}
where ${t_{0}} $ and $ b $ are constants. Thus, by substituting the above ansatz into equation \eqref{eq6} and integrating, we get
\begin{equation}
	\tau = \frac{t_{0}}{3b+1} \left( \frac{t}{t_{0}}\right)^{3b+1}\,,
\end{equation}
and by substituting this equation into equation \eqref{eq17}, we obtain
\begin{equation}
	a\left( \tau\right) = \left( \frac{\left(3b+1\right)  \tau}{t_{0}} \right)^{\frac{b}{3b+1}}\,.
\end{equation}
Expressed in terms of $\tau$, we consider the scale factor as
\begin{equation}\label{eq20}
	a\left(\tau\right)= \left( \frac{\tau}{\tau_{0}}\right)^{q}   \Longrightarrow \mathcal{H}= \frac{q}{\tau}\,,
\end{equation}
where $q$ and $\tau_{0}$ are constants. In this manner, the metric \eqref{eq7} takes the following form
\begin{equation}
		ds^{2} = -\left( \frac{\tau}{\tau_{0}} \right)^{-6q} d\tau^{2} + \left( \frac{\tau}{\tau_{0}} \right)^{2 q} \sum_{i=1}^3 \left( dx^{i}\right)^{2}\,.
\end{equation}
This relation shows that an accelerated universe happens for $\frac{1}{4}<q<\frac{1}{3}$. When $q=\frac{1}{3}$, the universe corresponds to a de Sitter cosmological evolution, and $q=\frac{2}{9}$ and $q=\frac{1}{5}$ correspond to dust and radiation dominated universes, respectively. By using the equation \eqref{eq20}, the Ricci scalar is given as
\begin{equation}
	R= \frac{-6q +30 q^{2}}{\tau_{0}^{2}} \left( \frac{\tau}{\tau_{0}}\right)^{6q- 2}\,.
\end{equation}
Then, equation \eqref{eq11} can be simplified as
\begin{equation}\label{eq23}
	\frac{6 q^{2}-2q}{\tau^{2}} f\left(\tau\right)+2(\frac{q}{\tau}) \dot{f} (\tau) + \ddot{f} \left(\tau\right)=0\,,
\end{equation}
so that the general solution of equation \eqref{eq23} is given by
\begin{equation}
	f\left( \tau\right)= C_{+} \tau^{\mu_{+}}+C_{-} \tau^{\mu_{-}}\,,
\end{equation}
where $C_{\pm}$ are integration constants and
\begin{equation}
	\mu_{\pm} = -q + \frac{1}{2} \pm \frac{\sqrt{-20 q^{2} + 4q+1}}{2}\,.
\end{equation}
Also, by substituting the above equation in equation \eqref{eq9}, the unimodular Lagrange multiplier $\lambda$ gets the following form:
\begin{equation}\label{eq26}
	\lambda = \mathcal{A}_{+} \tau^{\mu_{+}+6q-2}  + \mathcal{A}_{-} \tau^{\mu_{-} +6q-2}\,,
\end{equation}
where the parameters $ \mathcal{A}_{\pm} $ are defined as
\begin{equation}
	 \mathcal{A}\left( \tau\right)_{\pm} = \left[ 3 q^{2} + 3q \mu_{\pm} \right]  C_{\pm} \tau_{0}^{-6 q}.
\end{equation}
The Lagrange multiplier $\lambda$ given as equation \eqref{eq26} is a time-varying quantity, as expected to mimic the role of a cosmological constant from the unimodular viewpoint.

\section{ Non-minimal Unimodular Inflation in the Jordan Frame}\label{sec4}

The slow-roll parameters we have computed adhere to the conventional Planck definitions. In this respect, we have used the “Hubble Slow-Roll Approximation” (HSRA) introduced in \cite{liddle1994formalizing} for the Jordan frame since we discarded to consider a potential term in this frame. For the Einstein frame, since we are faced with an effective potential, we are allowed to use the “Potential Slow-Roll Approximation” (PSRA). On the other hand, as has been shown in \cite{liddle1994formalizing}, there is a hierarchy of HSR, but despite the potentially infinite number of slow-roll parameters, we require only a finite selection of them in any order. The first and second derivatives of the Hubble parameter are all we require to obtain first-order results in the slow-roll expansion. However, to go beyond this, we require more derivatives, necessitating further slow-roll parameters. Since in this paper we go beyond the first order slow-roll approximation, we considered more than the two slow-roll parameters, that is, four slow-roll parameters as will be seen later.

In this section, we start by presenting the basic equations in the non-minimal unimodular inflationary setup in the slow-roll approximation. We study cosmological inflation and calculate some important inflation parameters, such as the scalar spectral index and the tensor-to-scalar ratio. We consider the slow-roll approximations as \cite{fakir1990improvement}
\begin{equation}\label{eq28}
	\begin{split}
		\left|  \frac{\ddot{\phi}}{\dot{\phi}} \right|  \ll H\,,
		\qquad \left| \frac{\dot{\phi}}{\phi} \right|  \ll H\,,
		\qquad \left| \dot{H}\right| \ll H^{2}\,,
	\end{split}
\end{equation}
and we rewrite equation \eqref{eq28} as
\begin{equation}\label{eq29}
	\frac{d^2 \phi}{d\tau^{2}} \ll -2 \mathcal{H} \frac{d \phi}{d\tau}\,,
	\qquad \frac{d \phi}{d\tau}\ll \mathcal{H} \phi\,,
	\qquad \frac{d\mathcal{H} }{d\tau}\ll -2 \mathcal{H}^{2}\,.
\end{equation}
Under the slow-roll approximation, equation \eqref{eq14} is written as
\begin{equation}\label{eq30}
	\mathcal{H}^{2} = \frac{2 \lambda a^{-6}}{3 M_{P}^{2} \left( 1+\dfrac{\xi \phi^{2}}{M_{P}^{2}}\right)} \,,	
\end{equation}
and from equation \eqref{eq15} we obtain
\begin{equation}\label{eq31}
    -\frac{9}{2} M_{P}^{2} \mathcal{H}^{2} \left(1+\frac{\xi \phi^{2}}{M_{P}^{2}}\right) -5 \mathcal{H} \xi \phi \dot{\phi} + \lambda a^{-6} =0 \,.
\end{equation}
As we have explained in the previous part, we discarded the energy-momentum of the scalar field with the assumption that the NMC term, $\xi \phi^{2}R$, can play the same role. Using equation \eqref{eq3}, in the presence of the NMC between the scalar field and gravity, a contribution of this NMC term appears as
	 $$ \left( R_{\mu \nu}-\frac{1}{2} g_{\mu \nu} R\right) =- \frac{1}{f(\phi)} \left( \left(g_{\mu \nu} \Box - \nabla_{\mu} \nabla_{\nu}\right) f(\phi) +\lambda g_{\mu \nu} \right)$$
in the total energy-momentum of the problem. Our conjecture for the present model is that the energy-momentum contribution coming from the NMC is enough for forthcoming purposes. Therefore, by using equation $G_{\mu \nu}\equiv R_{\mu \nu}-\dfrac{1}{2} g_{\mu \nu} R = \frac{T^{(eff)}_{\mu \nu}}{M_{p}^{2}}$, we compute the effective energy-momentum as $ T_{\mu \nu}^{(eff)} \equiv - \frac{M_{P}^{2}}{f(\phi)} \left(\left(g_{\mu \nu} \Box - \nabla_{\mu} \nabla_{\nu}\right) f(\phi)+ \lambda g_{\mu \nu} \right)$. In this approach, we consider the continuity equation as
\begin{equation}\label{eq32}
	\dot{\rho}_{eff} + 3 \mathcal{H} \left( \rho_{eff} + P_{eff} \right) =0 \,,
\end{equation}
where $\rho_{eff}\left(\phi\right)$ and $P_{eff}(\phi)$ are the effective energy density and pressure of the generic field $\phi$ due to NMC, respectively. The effective energy density and pressure of $\phi$ can easily be obtained from the field equations \eqref{eq9} and \eqref{eq10}, which are as follows
\begin{equation}\label{eq33}
	\begin{split}
		& \rho_{eff}\left( \phi\right)  = - \frac{M_{P}^{2}}{f\left( \phi\right)} \left[3 \mathcal{H} \dot{f}\left( \phi\right) a^{6} - \lambda\left( \phi\right)\right]\,,\\
		& P_{eff}\left( \phi\right)  = \frac{M_{P}^2}{f(\phi)} \left[ \left( \ddot{f}(\phi)+ 5 \mathcal{H} \dot{f}(\phi) \right) a^{6} -\lambda(\phi) \right]\,.
	\end{split}
\end{equation}
From the above equations, we obtain
\begin{equation}\label{eq34}
	\frac{\dot{f}\left(\phi\right) }{f\left( \phi\right) }\left( 3\mathcal{H} \dot{f}\left( \phi\right) - \lambda\left(\phi\right)  a^{-6} \right) - 3 \dot{\mathcal{H}} \dot{f}\left( \phi\right)- 12 \mathcal{H}^{2} \dot{f}\left( \phi\right) + \lambda^{\prime}\left( \phi\right) \dot{\phi} a^{-6} =0\,,
\end{equation}
where
\begin{equation}\label{eq35}
	\lambda^{\prime} = \frac{d\lambda}{d\phi }\,.
\end{equation}
From now on, for simplicity we discard to write the explicit dependence of $\lambda$ and $\lambda^{\prime}$ to $\phi$. Now, the Klein-Gordon equation is given by
\begin{equation}\label{eq36}
	\xi \phi \ddot{\phi}+\xi\dot{\phi}^{2} + 6 \xi \phi \dot{\phi} \mathcal{H}+ \left(\frac{M_{P}^{2}  \lambda^{\prime} a^{-6}}{3 \xi \phi}\right)  \left[ 1+ \frac{\xi\phi^{2}}{M_{P}^{2}}\right]- \frac{4}{3} \lambda a^{-6}=0 \,.	
\end{equation}
Taking equation \eqref{eq36} and using the slow-roll conditions, we get
\begin{equation}\label{eq37}
	\dot{\phi} = \frac{2 \lambda a^{-6} }{9 \xi \phi \mathcal{H}} - \frac{M_{P}^2 \lambda^{\prime} a^{-6}}{18 \xi^{2} \phi^{2} \mathcal{H}} \left[ 1+ \frac{\xi \phi^2}{M_{P}^2}\right]\,.
\end{equation}
Furthermore, by using the equations \eqref{eq30} and \eqref{eq37}, we obtain
\begin{equation}
	\dot{\mathcal{H}}= \frac{\lambda^{\prime} a^{-6}}{6 \xi \phi} - \frac{M_{P}^{2} \left( 1+\dfrac{\xi \phi^{2}}{M_{P}^{2}}\right) \lambda^{\prime 2} a^{-6}}{36 \lambda \xi^{2} \phi^{2}}
	-\frac{20 \lambda a^{-6}}{9 M_{P}^{2} \left(1+\dfrac{\xi \phi^{2}}{M_{P}^{2}}\right)}\,.
\end{equation}

\subsection{Non-minimal Unimodular Inflation and the Slow-roll Parameters}

In the Jordan frame, we define the slow-roll parameters as follows \cite{liddle1994formalizing, nojiri2017modified}:
\begin{equation}\label{eq39}
	\epsilon_{1} = - \frac{\dot H}{H^{2}}\,, \quad
\epsilon_{2} =  \frac{\ddot{\phi}}{H \dot{\phi} }\,, \quad
\epsilon_{3} = \frac{\dot{f}\left(\phi\right) }{2 H f\left(\phi\right) }\,, \quad
\epsilon_{4} =\frac{\dot{E}}{2 H E}\,,
\end{equation}
where, by definition
\begin{equation}\label{eq40}
	E\equiv f\left(\phi\right)+ \frac{3 \dot{f}^{2}\left(\phi\right)}{2 \kappa^{2} \dot{\phi}^{2}}\,.
\end{equation}
Moreover, the scalar spectral index and the tensor-to-scalar ratio in terms of the slow-roll parameters are represented respectively as
\begin{equation}\label{eq41}
	n_{s} \simeq 1- 4 \epsilon_{1} - 2 \epsilon_{2} +2\epsilon_{3} -2 \epsilon_{4}\,,
\end{equation}
\begin{equation}\label{eq42}
r= 8 \kappa^{2} \frac{Q_{s}}{f\left( \phi\right)}\,,
\end{equation}
where the parameter $Q_{s}$ is given by
\begin{equation}\label{eq43}
Q_{s}= \frac{E \dot{\phi^{2}} }{f\left( \phi\right)  H^{2} (1 +\epsilon_{3} )^{2}}\,.
\end{equation}
Note that the slow-roll parameters satisfy the slow-roll condition $\epsilon_{i} \ll 1, i=1,...4 $. Also, the e-folds number is defined as follows
\begin{equation}
	N= \int_{t_{hc}}^{t_{e}} H \, dt = \int_{\phi_{hc}}^{\phi_{e}} \frac{H}{\dot{\phi}} \, d\phi\,,
\end{equation}
where ${t_{hc}}$ is the horizon crossing time, and $\phi_{hc}$ is the value of the scalar field where the CMB scale crosses the horizon. $t_{e}$ is the time that inflation ends, and $\phi_{e}$ is the final value of the inflaton, $\phi$, during the slow roll era. Hence, we can rewrite the slow-roll parameters \eqref{eq39} as
\begin{equation}
	\epsilon_{1}= -3 - \frac{\dot{\mathcal{H}}}{\mathcal{H}^{2}}\,, \quad
	\epsilon_{2}= 3+\frac{\ddot{\phi}}{\mathcal{H} \dot{\phi}}\,, \quad
	\epsilon_{3}= \frac{\dot{f}\left( \phi\right)}{2 \mathcal{H} f\left( \phi\right) }\,, \quad
	\epsilon_{4}= \frac{\dot{E}}{2 \mathcal{H} E}\,.
\end{equation}
Therefore, the function $E$ is equal to
\begin{equation}
	E= 1 + \left(1 + 6 \xi \right)   \frac{\xi\phi^{2}}{M_{P}^{2}}\,,
\end{equation}
and
\begin{equation}
	\dot{E}=  \frac{2 \left( 1 + 6 \xi \right) }{M_{P}^{2}} \dot{f}\left( \phi\right) \,.
\end{equation}
To continue, by combining equations \eqref{eq42} and \eqref{eq43}  and the above equation, we obtain
\begin{equation}
	r= \frac{2 M_{P}^{2} E \dot{\phi^{2}} }{f^{2}\left( \phi\right)\mathcal{H}^{2} (1+\epsilon_{3})^{2} }\,.
\end{equation}
Thus, we compute the slow-roll parameters as
\begin{align}\label{eq49}
\epsilon_{1} = &   \frac{ \left( M_{P}^{2} \left(1+  \dfrac{\xi \phi^{2}}{M_{P}^{2}} \right)  \lambda^{\prime} - 2 \xi \phi  \lambda \right) \left( M_{P}^{2} \left(1+  \dfrac{\xi \phi^{2}}{M_{P}^{2}} \right)  \lambda^{\prime}  -4 \xi \phi \lambda \right)  }{24 \xi^{2} \phi^{2}\lambda^{2} }\,,\\
	 \begin{split}\label{eq50}
\epsilon_{2} = & \left[  \frac{1}{ 72 \lambda^{2} \left( M_{P}^{2}\left(1+  \dfrac{\xi \phi^{2}}{M_{P}^{2}} \right)  \lambda^{\prime} - 4 \lambda \xi \phi \right) \xi^{2}\phi^{3}} \right]   \Biggl\{ -6 \phi\lambda \lambda^{\prime\prime} \left[  M_{P}^{2}\left(  1+  \frac{\xi\phi^{2}}{M_{P}^{2}} \right) \lambda^{\prime} - 4 \lambda \xi \phi \right] \left( M_{P}^{4} \left(  1+  \frac{\xi\phi^{2}}{M_{P}^{2}} \right)^{2} \right) \\
	       	 & + 3 \phi {M_{P}^{6}} \left(  1+ \frac{ \xi \phi^{2}}{M_{P}^{2}} \right)^{3}  \lambda^{\prime 3} + 12 \lambda \lambda^{\prime 2} \left( \frac{17 \xi \phi^{2}}{6} + M_{P}^{2} \right) M_{P}^{4} \left(  1+ \frac{ \xi\phi^{2}}{M_{P}^{2}} \right)^{2}  + 96 \left( \xi \left( M_{P}^{2}-1 \right)\phi^{2}+M_{P}^{4} \right) \lambda^{3} \xi^{2} \phi^{2}\\
	    	 & -72 \xi \phi  \lambda^{2} \lambda^{\prime} \left( 148 \xi \phi^{2} +M_{P}^{2} \right) M_{P}^{2} \left(  1+ \frac{ \xi\phi^{2}}{M_{P}^{2}} \right)  \Biggr\}\,,
	 \end{split}\\
\epsilon_{3} = & \frac{1}{3}- \frac{ M_{P}^{2} \left(  1+  \dfrac{\xi\phi^{2}}{M_{P}^{2}} \right) \lambda^{\prime}}{ 12 \xi \phi \lambda}\,, \\
\epsilon_{4} = & - \dfrac{\left( \dfrac{1}{6}+ \xi \right) \left( M_{P}^{2} \left(1+ \dfrac{\xi\phi^{2}}{M_{P}^{2}} \right)  \lambda^{\prime} -4 \xi \phi \lambda \right) M_{P}^{2} \left(1+ \dfrac{\xi\phi^{2}}{M_{P}^{2}} \right)   }{2 \xi \phi \lambda \left(6 \xi^{2} \phi^{2} + \xi \phi^{2} +  M_{P}^{2} \right)}\,,  \label{eq52}
	\end{align}
and the number of e-folds is
\begin{equation}
	N =  - \int_{\phi_{hc}}^{\phi_{e}} \frac{12 \lambda \xi^{2} \phi^{2} }{\left( \xi \phi^{2} + M_{P}^{2}\right) \left( \left( \xi \phi^{2} + M_{P}^{2}\right) \lambda^{\prime}- 4 \lambda  \xi \phi \right)}\, d\phi \,,
\end{equation}
where as before, $ \lambda^{\prime}= \dfrac{d\lambda}{d\phi}$\,.

\subsection{A Power Law Lagrange Multiplier}

Now, we consider $\lambda$ as
\begin{equation}\label{eq53}
	\lambda(\phi)= A \phi^{n}\,,
\end{equation}
where $A$ is a constant and $n$ is an arbitrary power that could be constrained via observational data in the same way as the power law potentials in Planck2018.
Since in our case $\lambda$ is not a constant, there is essentially a non-conservation of energy momentum in this setup. In equation \eqref{eq32} we have introduced effective density and pressure that contain the effects of the generic non-minimal scalar field $ \phi $ and the unimodular parameter $\lambda $. So, the effective energy-momentum could be conserved, but not the energy-momentum of the non-minimally coupled scalar field with variable $\lambda$ as \eqref{eq53}. As equation \eqref{eq35} shows explicitly, $\lambda$ is not constant in our setup, and therefore this differs from the results of GR with a non-minimally coupled scalar field. In fact, if instead of defining effective quantities as equation \eqref{eq32}, we were to consider the contribution of $ \phi $ and $\lambda$ as they appear, then we were faced with nonconservation of the energy-momentum of the field obviously. 

Let us consider $\beta^{2} \equiv \xi \phi_{end}^{2}/ M_{P}^{2} $. Inflation ends when, for instance, $ \epsilon_{1} \simeq 1$. Then, by using equations \eqref{eq49} and \eqref{eq53} we find $\beta$ as
\begin{equation}
	\beta = \pm \frac{\sqrt{-(n+2) n}}{n+2}\,,\quad  \pm \frac{\sqrt{-(n-8) n}}{n-8}\,.
\end{equation}
If we write $ m^{2} \equiv \xi \phi_{hc}^{2} /M_{P}^{2}$, the slow-roll parameters $\epsilon_{i=1...4}$ are obtained as follows
\begin{equation}
	\begin{split}
		&  \epsilon_{1}= \frac{\left( \left( n-2 \right) m^{2}+n\right) \left(  \left( n-4 \right) m^{2}+n \right)}{24 m^{4}}\,,	\\
		& \begin{split}
			\epsilon_{2}= & \frac{1}{72 m^{4} ((n-4) m^{2}+n)} \biggl\{\left(-3 n^{3}+64 n^{2}-10680 n \right)m^{6} +\left(-9 n^{3}+146 n^{2}-10776 n +96\right)m^{4}\\
			& +\left(-9 n^{3}+100 n^{2}-96 n\right)  m^{2}-3 n^{3}+18 n^{2} \biggr\}\,,
		\end{split}\\
		& \epsilon_{3}=\frac{\left(-n+4\right) m^{2}-n}{12 m^{2}}\,, \\
		&  \epsilon_{4} = -\frac{6 \left( 1/6 + \xi \right) \left( m^{2}+1\right) \left( \left( n-4\right)  m^{2} + n \right)}{\left(12+72 \xi \right)  m^{4}+12 m^{2}}\,,
	\end{split}
\end{equation}
where $\epsilon_{i=1...4} \ll 1 $ must be fulfilled in the slow-roll regime. It is clear by these expressions that the inflation's observable quantities, including the spectral index and tensor-to-scalar ratio, are dependent on $\xi$ in this setup.
Ultimately, the number of e-folds of the non-minimal unimodular inflation in the Jordan frame derives as follows
\begin{equation}
\begin{split}
		N & = \left[ \frac{3 \ln\left( \xi \phi^{2} + M_{P}^{2} \right) }{2} - \frac{3n \ln\left( \xi \phi^{2} \left( n-4 \right) \right)+ n M_{P}^{2} }{2n-8}\right]\Bigg |_{\phi_{hc}}^{\phi_{e}}\\
	& = \left( \frac{3}{2}\right) \ln\left( \frac{m^{2} + 1}{\beta^{2}+1}\right) + \left( \frac{3n}{2n-8} \right) \ln \left( \frac{\beta^{2} (n-4)+n}{m^{2}(n-4) +n} \right)\,.
\end{split}
\end{equation}
Accordingly, we can express the spectral index $n_{s}$ versus the NMC parameter, $\xi $, as follows
\begin{equation}
	\begin{split}
		n_{s}= &\bigg[ \frac{1}{216 m^{4} \left( 1/6 + \left( 1/6+\xi \right) m^{2}\right) \left( \left( n-4\right) m^{2} + n \right) } \bigg] \biggl\{ -18\left( 1/6+\xi \right)  \left( n^{3} + \frac{4}{3} n^{2} -3508 n -16 \right) m^{8} \\
		&+ \left( -48 + \left( -54 \xi - 12 \right) n^{3} + \left( -120 \xi -30 \right) n^{2} +  \left( 63432 \xi + 21144 \right) n \right)  m^{6} \\
		& + \left( -96+ \left( -54 \xi -18 \right) n^{3}
		 + \left( -168 \xi - 66 \right) n^{2} + \left( 288 \xi +10716 \right) n \right) m^{4} \\
		& + \left( \left( -18 \xi -12 \right) n^{3} +\left( -72 \xi -54 \right) n^{2} + 96 n \right) m^{2}  -3 n^{3} -18 n^{2}  \biggr\}\,.
	\end{split}
\end{equation}
Likewise, the tensor-to-scalar ratio $r$ is obtained as follows
\begin{equation}
	r= \frac{48 \left[  1/6 + \left(  \xi + 1/6 \right) m^{2} \right]  \left[  \left( n-4 \right)m^{2} + n \right]^{2} }{\left[  \left(  n-16\right) m^{2} + n \right]^{2} \xi m^{2}}\,.
\end{equation}

\section{Field Equations of Non-minimal Unimodular Gravity in the Einstein Frame}\label{sec5}

In this section, we consider the action  \eqref{eq13} and apply a Weyl or conformal transformation (see also \cite{kaiser1995primordial}) as follows
\begin{equation}
{g}_{\mu\nu} \longrightarrow	\tilde{g}_{\mu\nu}=\Omega^{2} g_{\mu \nu}\,,   \qquad
 \sqrt{-g}=\Omega^{-4} \sqrt{-\tilde{g}}\,,   \qquad
 \Omega^{2}=f\left(\phi\right) \,,
\end{equation}
where $\Omega = \Omega\left( \phi\right)$  is the conformal factor. Then, we get the action in the Einstein frame as
\begin{equation}\label{eq61}
	S_{E}= \int d^{4}x  \left \lbrace  \sqrt{ -\tilde{g}} \left( \frac{\tilde{R}}{2\kappa^{2}} - \frac{1}{2} \tilde{g}^{\mu \nu} \nabla_{\mu}\varphi \nabla_{\nu}\varphi \right) - U\left( \varphi\right)  \right \rbrace\,.
\end{equation}
Therefore, we have
\begin{equation}
  U(\varphi)=\frac{R \Omega^{2}- R f\left( \phi\right)}{ \Omega^{4} } +  \frac{ 2 \lambda -\frac{2 \lambda}{\sqrt{-g}} }{2 \kappa^{2} \Omega^{4}}\,,
\end{equation}
 and we define a new scalar field $\varphi$ as
\begin{equation}\label{eq63}
   \kappa \varphi \equiv \sqrt{\frac{3}{2} }\ln \Omega^{2}\,, \qquad
   \qquad
   \Omega^{2}= e^{\sqrt{2/3} \kappa \varphi} \,.
\end{equation}
Also, the Ricci scalar in the Einstein frame is as follows
\begin{equation}
	\tilde{R} = \Omega^{-2} \left[ R -6 \Box \left( \ln{\Omega} \right) - 6 \nabla_{a} \left( \ln{\Omega} \right) \nabla^{a} \left( \ln{\Omega} \right) \right]\,.
\end{equation}
In light of these definitions, we can rewrite the action \eqref{eq61} as
\begin{equation}\label{eq66}
	S_{E}= \int d^{4}x  \left \lbrace  \sqrt{-\tilde{g}} \left[\frac{\tilde{R}}{2\kappa^{2}} - \frac{1}{2} \tilde{g}^{\mu \nu} \nabla_{\mu}\varphi \nabla_{\nu}\varphi -U(\varphi) \right]-2 \tilde{\lambda} \left(e^{-2\sqrt{2/3}\kappa\varphi} \sqrt{-\tilde{g}} -1 \right) \right \rbrace \,,
\end{equation}
where $\tilde{\lambda} = \dfrac{\lambda}{2 \kappa^{2}}$.
If we set $\Omega^{2}= \frac{2}{M_{P}}f\left( \phi\right)  $, then $U\left( \varphi\right)=0$. In this regard, the variation of the action \eqref{eq66} with respect to the Lagrange multiplier $ \tilde{\lambda}$ leads to the constraint equation as
\begin{equation}
	\sqrt{- \tilde{g}}=  e^{2\sqrt{2/3}\kappa \varphi}\,,
\end{equation}
and consequently, the determinant of the metric $\tilde{g}_{\mu \nu} $ is not a constant \cite{saez2016analyzing}. By varying the action equation \eqref{eq66} with respect to the metric, the field equations are obtained as follows
\begin{equation}
	\tilde{R}_{\mu\nu} -\frac{1}{2} \tilde{g}_{\mu \nu} \tilde{R} = \kappa^{2} \left[ \nabla_{\mu}\varphi \nabla_{\nu}\varphi -\frac{1}{2}\tilde{g}_{\mu \nu} \nabla^{\rho}\varphi \nabla_{\rho}\varphi - 2 \tilde{\lambda} \tilde{g}_{\mu \nu}  e^{-2\sqrt{2/3}\kappa \varphi} \right]\,,
\end{equation}
and
\begin{equation}
\tilde{G}_{\mu \nu} = \kappa^{2} \tilde{T}_{\mu \nu}\,,
\end{equation}
where $\tilde{R}_{\mu\nu} $ is the Ricci tensor, $ \tilde{R} $ is Ricci scalar and $\tilde{T}_{\mu \nu}$ is the matter energy-momentum tensor as follows
\begin{equation}\label{eq71}
	\tilde{T}_{\mu \nu} = \nabla_{\mu}\varphi \nabla_{\nu}\varphi - \frac{1}{2} \tilde{g}_{\mu \nu} \nabla^{\rho}\varphi \nabla_{\rho}\varphi - 2 \tilde{\lambda} \tilde{g}_{\mu \nu} e^{-2\sqrt{2/3}\kappa\varphi}\,,
\end{equation}
In the next step, we compute the scalar field equation of motion by varying the action \eqref{eq66} with respect to the field $\varphi$ (that is, $\delta S_{J}/ \delta \varphi = 0$),
 \begin{equation}\label{eq72}
 	\tilde{\Box} \varphi + 4 \kappa \tilde{\lambda} \sqrt{\frac{2}{3}} e^{-2\sqrt{2/3} \kappa \varphi}=0 \,.
 \end{equation}
By using $\nabla_{\mu}\tilde{G}^{\mu}_{_{\nu}} =0 $ and taking the divergence of the field equation \eqref{eq72}, the matter-energy conservation $\nabla_{\mu}\tilde{T}^{\mu}_{_{\nu}} =0 $ becomes
\begin{equation}\label{eq73}
	\nabla_{\mu}\tilde{T}^{\mu}_{_{\nu}}= \left( \tilde{\Box}\varphi + 4 \kappa \tilde{\lambda} \sqrt{\frac{2}{3}} e^{-2\sqrt{2/3}\kappa\varphi} \right)\partial_{\nu} \varphi  - 2 e^{-2\sqrt{2/3}\kappa\varphi}\partial_{\nu} \tilde{\lambda} =0 \,.
\end{equation}
In equation \eqref{eq73}, the first term vanishes; thus we get
\begin{equation}
      \partial_{\nu}\tilde{\lambda } = 0\,,    \longrightarrow \tilde{\lambda} = \tilde{\lambda}_{0}\,.
\end{equation}
Hence, the energy-momentum tensor equation \eqref{eq71} is equal to
\begin{equation}
	\tilde{T}_{\mu \nu} = \nabla_{\mu}\varphi \nabla_{\nu}\varphi-\frac{1}{2} \tilde{g}_{\mu \nu}  \nabla^{\rho}\varphi \nabla_{\rho}\varphi - 2 \lambda_{0} \tilde{g}_{\mu \nu} e^{-2\sqrt{2/3}\kappa\varphi}\,.
\end{equation}
In this setup, by considering $ \tilde{R}_{\mu\nu} -\frac{1}{2} \tilde{g}_{\mu \nu} R=\kappa^{2} \tilde{T}_{\mu \nu}  $, we redefine the action equation \eqref{eq61} as
\begin{equation}
	S_{E}= \int d^{4} x  \left \lbrace  \sqrt{-\tilde{g}} \left(   \frac{\tilde{R}}{2\kappa^{2}} - \frac{1}{2} \tilde{g}^{\mu \nu} \nabla_{\mu}\varphi \nabla_{\nu}\varphi \right)- U_{eff}\left( \varphi\right) \right \rbrace\,,
\end{equation}
where we defined the effective potential as follows,
\begin{equation}\label{eq77}
	U_{eff}\left( \varphi\right)= 2 \tilde\lambda_{0} e^{-2\sqrt{2/3}\kappa \varphi}\,.
\end{equation}
This action is apparently minimal.

\subsection{Non-minimal Coupling of the Scalar Field, $f\left(\phi\right) $}

Here to study the cosmological solutions, we assume a flat unimodular FRW metric in the Einstein frame as follows
\begin{equation}
	d\tilde{s}^{2}=\Omega^{2} ds^{2}= -\tilde{a}^{-6} (\tilde{\tau}) d\tilde{\tau}^{2} + \tilde{a}^{2} (\tilde{\tau}) dx_{i} dx^{i} \,.
\end{equation}
We can see that $\tilde{a} \equiv \Omega a = \sqrt{f} a $ and $ d\tilde{\tau} \equiv \Omega d\tau = \sqrt{f} d\tau $.
Let us assume the Higgs field coupling $f(\phi)$ in the following way
\begin{equation}
	f\left(\phi\right)= \frac{1}{2} M_{P}^{2} \left( 1+\frac{\xi \phi^{2}}{M_{P}^{2}}\right)\,.
\end{equation}
Now, we consider the conformal transformation, which redefines the \textquotedblleft potential", leading to
\begin{equation}\label{eq80}
	\Omega^{2} = \frac{2 f\left( \phi\right) }{M_{P}^{2}} =\left( 1+ \frac{\xi \phi^{2}}{M_{P}^{2}}\right)\,,
\qquad
U_{eff}\left(\varphi\right) \equiv \frac{ 2\tilde\lambda_{0}  }{\Omega^{4}}= \frac{2\tilde\lambda_{0} }{ \left(1+ \dfrac{\xi \phi^{2}}{M_{P}^{2}}\right)^{2}}\,,	
\end{equation}
and the relation between $ \phi $ and $ \varphi $ is written as
\begin{equation}\label{eq81}
\frac{d\varphi}{d\phi}\equiv {M_{P}} \sqrt{\frac{f\left( \phi\right)+ 3 \left(f^{\prime}\left( \phi\right) \right)^{2}}{2 f^{2}\left( \phi\right) }} = \sqrt{\frac{1+ \left( \frac{\xi \phi^{2}}{M_{P}^{2}}\right)  \left( 1+6 \xi \right) }{\left( 1+ \frac{\xi \phi^{2}}{M_{P}^{2}} \right)^{2}}}\,.
\end{equation}
Therefore, the Friedmann equations in the Einstein frame take the following form
\begin{equation}
	\begin{split}
		&2 \dot{\tilde{\mathcal{H}}} + 9 \tilde{\mathcal{H}}^{2}= -  \frac{1}{M_{P}^{2}} \left(  \frac{1}{2}  \dot{\varphi}^{2} - U_{eff}\left( \varphi\right) \right) \,, \\
		& 3 \tilde{\mathcal{H}}^{2} = \frac{1}{M_{P}^{2}} \left(\frac{1}{2}  \dot{\varphi}^{2} +  U_{eff}\left( \varphi\right) \right) \,,  \\	
		& \ddot{\varphi} + 3 \mathcal{H} \dot{\varphi}- U_{eff}^\prime\left( \varphi\right) =0 \,,
	\end{split}
\end{equation}
where overdot as usual indicates derivative with respect to the time parameter $\tau$ and prime denotes $d/ d\varphi$.

\subsection{Slow-roll Field Equations}

By using the \emph{potential-slow-roll approximation}, $ U_{eff}\left( \varphi\right)  \gg \dot{\varphi}^{2} $ and $ \ddot{\varphi} \ll 3  \tilde{\mathcal{H}} \dot{\varphi} $, the field equations during the slow-roll stage lead to
\begin{equation}
	3 \tilde{\mathcal{H}}^{2} \approx \frac{U_{eff}\left( \varphi\right)}{M_{P}}\,,
\end{equation}
\begin{equation}
	 \dot{\varphi} \approx \frac{U_{eff}^\prime\left(\varphi\right)}{3 \tilde{\mathcal{H}}}\,.
\end{equation}
We study these conditions for the first and second orders in the potential-slow-roll approximation parameters. So we calculate three of the potential-slow-roll parameters as follows
\begin{equation}\label{eq85}
\begin{split}
	&	\epsilon = \frac{M_{P}^{2}}{2} \left( \frac{U_{eff}^{\prime}\left(\varphi\right) }{U_{eff}\left(\varphi\right)} \right)^{2}\,, \\
	&	\eta = M_{P}^{2} \left(  \frac{U_{eff}^{\prime\prime}\left( \varphi\right) }{U_{eff}(\varphi)} \right)\,,  \\
	& 	\zeta = M_{P}^{2} \left( \frac{U_{eff}^{\prime}\left(\varphi\right)  U_{eff}^{\prime\prime\prime}\left( \varphi\right)}{U_{eff}^{2}\left(\varphi\right) } \right)^{\frac{1}{2}}\,.
\end{split}
	\end{equation}
By using the equations \eqref{eq80} and \eqref{eq81}, and the potential-slow-roll parameters as equation \eqref{eq85}, we find
\begin{equation}\label{eq86}
	\begin{split}
		& \epsilon =\frac{8 \xi^{2} \phi^{2} }{ M_{P}^{2} + \left( 1+ 6 \xi \right) \xi \phi^{2}}\,, \\
		& \eta= -\frac{ 4 \xi \left[ -24  \xi^{3} \phi^{4} -4 \xi^{2} \phi^{4} -3 \xi \phi^{2} M_{P}^{2} + M_{P}^{4} \right]  }{  \left[ 6 \xi^{2} \phi^{2} + \xi \phi^{2} + M_{P}^{2}  \right]^{2}}\,, \\
		& \zeta = 4 M_{P}^{2} \sqrt{2} \sqrt{\dfrac{\splitdfrac{288 \xi^{8} \phi^{8} + 96 \xi^{7} \phi^{8} + 54 \xi^{6} \phi^{6} \left(  M_{P}^{2} +  4 \phi^{2}/ 27  \right)  }{-54 \phi^{4} M_{P}^{2} \xi^{5}  \left( M_{P}^{2}- \phi^{2}/ 6  \right)- 12 \phi^{2} M_{P}^{4} \xi^{4} \left( M_{P}^{2} + \phi^{2}/ 2  \right) - 7 M_{P}^{6} \xi^{3} \phi^{2}}}{ M_{P}^{4} \left[ 6 \xi^{2} \phi^{2} + \xi \phi^{2} +  M_{P}^{2} \right]^{4}}}\,.
	\end{split}
\end{equation}
Likewise, in the Jordan frame, it is clear by these expressions that the Higgs inflation's observable quantities, including the spectral index and tensor-to-scalar ratio, are dependent on $\xi$.
In the Einstein frame, the end of inflation occurs at $\epsilon = 1$. If we assume $\beta^{2}\equiv \xi \phi_{end}^{2}/ M_{P}^{2} $, we find
\begin{equation}\label{eq87}
	\beta= \pm \frac{1}{\sqrt{2 \xi -1}}\,,
\end{equation}
which exclude $\xi=1/2$ in order to be well-defined.
We use the definition of the number of e-folds, to find
\begin{equation}\label{eq88}
	N = \int_{\varphi_{hc}}^{\varphi_{e}}    \frac{\tilde{H}}{\dot{\varphi}} \, d\varphi =  \frac{1}{M_{P}^{2}} \int_{\phi_{hc}}^{\phi_{e}} \frac{U_{eff}(\varphi)}{dU_{eff}(\varphi)/d\phi} \left( \frac{d\varphi}{d\phi} \right)^{2}  \, d\phi = \left[ \frac{-3 \xi \ln{(\xi \phi^{2} + M_{P}^{2})}-\ln{\phi} }{4\xi}\right] \bigg |_{\phi_{hc}}^{\phi_{e}}\,.
\end{equation}
Here, we have the first and second orders spectral index  $n_{s}$ and the tensor-to-scalar ratio $ r $ \cite{liddle1994formalizing}, where the first order quantities are geiven by
\begin{equation}
	\begin{split}
		& n_{s} = 1- 6 \epsilon + 2 \eta \,,\\
		& r= 16 \epsilon \,,
	\end{split}
\end{equation}
and the second order ones are given by \cite{liddle1994formalizing}
\begin{equation}\label{eq89}
\begin{split}
	&	n_{s} = 1- 6 \epsilon + 2 \eta + \dfrac{1}{3} \left( 44 - 18 c \right) \epsilon^{2} + \left( 4 c - 14  \right)  \epsilon  \eta + \frac{2}{3} \eta^{2} + \frac{1}{6} \left( 13 - 3 c \right)  \zeta^{2}+... \,, \\
	&  r= \left( \frac{25}{2}\right) \epsilon \left[1+ 2 \left( c- \frac{1}{3} \right) \left( 2 \epsilon - \eta \right) +...\right]\,,
\end{split}
\end{equation}
where by definition $ c \equiv 4(\ln2 + \gamma)\simeq 5.081 $ and $\gamma \simeq 0.577$ is the Euler's constant. Following the Jordan frame, if we write $ m^{2}\equiv \xi \phi_{hc}^{2} /M_{P}^{2}$, then equations \eqref{eq86} become as
\begin{equation}
	\begin{split}
		& \epsilon =\frac{8 \xi m^{2} }{ 1 + \left( 1+ 6 \xi \right) m^{2}}\,, \\
		& \eta= -\frac{ 4 \xi \left[-24 \xi m^{4}  -4  m^{4} -3 m^{2} + 1 \right] }{  \left[   6 \xi m^{2} + m^{2} +1  \right]^{2}}\,, \\
		&  \zeta = 4 \sqrt{2} \sqrt{\dfrac{\splitdfrac{m^{2} \xi^{2} \big(  288 m^{6} \xi^{2} + 96 m^{6} \xi + 54 m^{4} \xi- 54 m^{2} \xi}{+ 8 m^{6} + 9 m^{4}  -6 m^{2}- 12 \xi- 7 \big) }}{\left[  6\xi m^{2} + m^{2} +  1 \right]^{4}}}\,.
	\end{split}
\end{equation}
To first order, we calculate the slow-roll parameters in terms of $m$ and $ \xi $ as
\begin{equation}
	n_{s} = \frac{\left( -60 \xi^{2} - 4 \xi +1\right) m^{4} + \left( -12 \xi + 2 \right)m^{2} - 8 \xi +1 }{\left(6 \xi m^{2} + m^{2} +1  \right)^{2} }\,,
\end{equation}
and
\begin{equation}
	r= \frac{128 \xi m^{2}}{\left( 6 \xi +1 \right) m^{2} +1 }\,.	
\end{equation}
In the same way, to the second order, the spectral index in this frame can be obtained as follows
\begin{equation}
\begin{split}
n_{s}= & 1- \frac{48 \xi  m^{2}}{\left( 1+ 6 \xi\right)  m^{2} +1 }- \frac{8 \xi \left( -24 \xi m^{4}  -4  m^{4} -3 m^{2} + 1\right) }{\left( 6 \xi  m^{2} +  m^{2} +1 \right) }- \frac{1012.43 \xi^{2}  m^{4}}{\left( \left( 1+ 6 \xi \right)  m^{2} + 1 \right)^{2} }\\
	  & - \frac{202.36 \xi^{2}  m^{2} \left( -24 \xi m^{4}  -4  m^{4} -3 m^{2} + 1 \right) }{\left( 6 \xi  m^{2} +  m^{2} +1 \right)^{2} \left( \left( 1+ 6 \xi\right)  m^{2} +1  \right) } + \frac{32 \xi^{2} \left( -24 \xi m^{4}  -4  m^{4} -3 m^{2} + 1 \right)^{2}}{3 \left( 6 \xi  m^{2} +  m^{2} +1 \right)^{4}}\\
	  & - \frac{11.96 m^{2} \xi^{2} \left( 288\xi^{2} m^{6} +96 \xi m^{6} +54 \xi m^{4} - 54 \xi m^{2}  + 8 m^{6}+ 9 m^{4} -6 m^{2} -12 \xi -7 \right) }{\left( 6 \xi  m^{2} +  m^{2} +1 \right)^{4}}\,,
\end{split}
\end{equation}
and the second order tensor-to-scalar ratio in the Einstein frame becomes as follows
\begin{equation}
	r= \frac{100 \xi m^{2} \left(36 \xi^{2}  m^{4} + 12 \xi  m^{4} + m^{4} + 49.98  \xi  m^{2} + 2 m^{2}+ 37.98 \xi +1 \right) }{\left( 6 \xi  m^{2} +  m^{2} +1 \right)^{3}}\,.
\end{equation}
Finally, the e-folds number $N$ is calculated as
\begin{equation}
	N=\frac{3}{4} \ln\left(\frac{m^{2}+1}{\beta^{2}+1} \right) + \frac{1}{4\xi} \ln\left( \frac{m}{\beta} \right) \,.
\end{equation}

After completion of the mathematical structure of the our conjectured non-minimal unimodular model, we are now in the position that we can perform some numerical analysis on the parameters space of the model in order to see viability and feasibility of the model and to find some novel severe constraints on the model parameters in confrontation with recent observational data.

\section{Analysis of Parameter Space and Comparison with Observation}\label{sec6}

In this section, we study numerically the parameter space of the model consisting of $ m $, $ \xi $, and $n$ in the two frames, and the results are compared with observational data. In our analysis, we check consistency with observational data and obtain some novel and severe constraints on the parameter space of the model. The predictions of our model are consistent with both Planck2018 TT, TE, EE+lowE+lensing+BAO+BK\textbf{14}, based on the $ \varLambda CDM+r+\frac{dn_{s}}{dn_{k}} $ model, with the value of the scalar spectral index as $n_{s}=0.9658\pm 0.0038$ and upper limit on the tensor-to-scalar ratio $r<0.072$, and also Planck2018 TT, TE, EE+lowE+lensing+BAO+BK\textbf{18} upper limit on the tensor-to-scalar ratio tighten to $ r<0.036$ at $ 68\% $ and $ 95\% $ confidence levels. The forthcoming plots present the results of our study. Our goal is to find some novel and severe constraints on the value of the NMC parameter $\xi $. For this purpose, we used the observational values of the $n_s$ and $r$ from the mentioned joint data sets and their confidence levels. Each of these two observables have some observational limits. To proceed, due to the wide parameter space of the model, we fixed the values of the parameters $m$ and $n$ at some appropriate values. Since we were to have the consistency of the model with observational data, the appropriateness here means those values of the mentioned two parameters ($m$ and $n$) that save the model in the confidence levels of the used data sets.

\subsection{CMB Constraints}
\begin{figure}[H]
	\centering
	\includegraphics[height= 7cm, width=14cm]{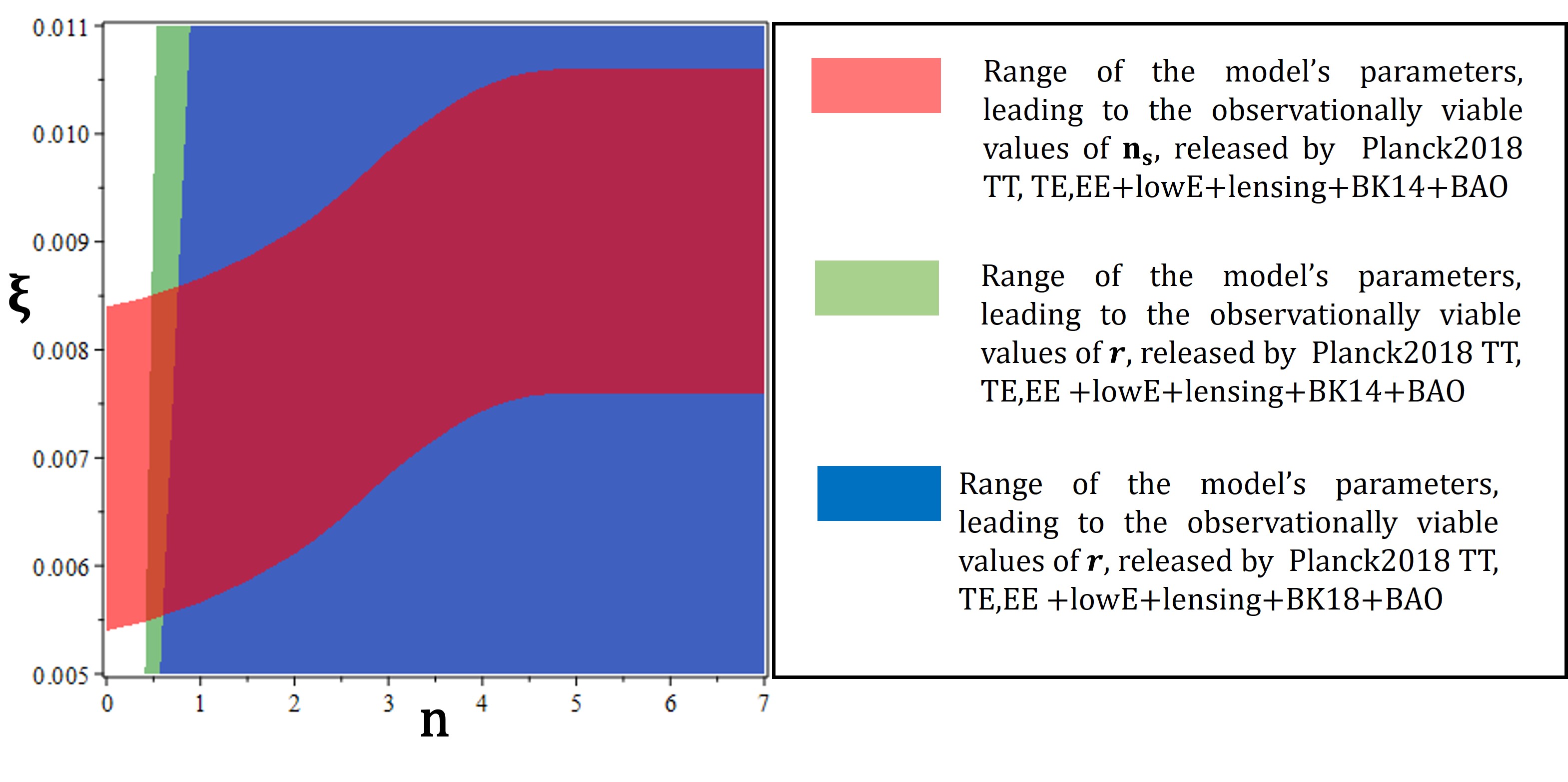}
	\caption{Ranges of parameters $ \xi $ and $n $ for $\lambda=A\phi^{n} $ model in the Jordan frame with constant $ m= \pm 0.505$. In this case, the constraints on the NMC parameter is $0.0055 \leq \xi \leq 0.0105$, where $n\geq0.525$.}\label{figure.1}
\end{figure}

\begin{figure}[H]
	\centering
	\includegraphics[height= 7cm, width=14cm]{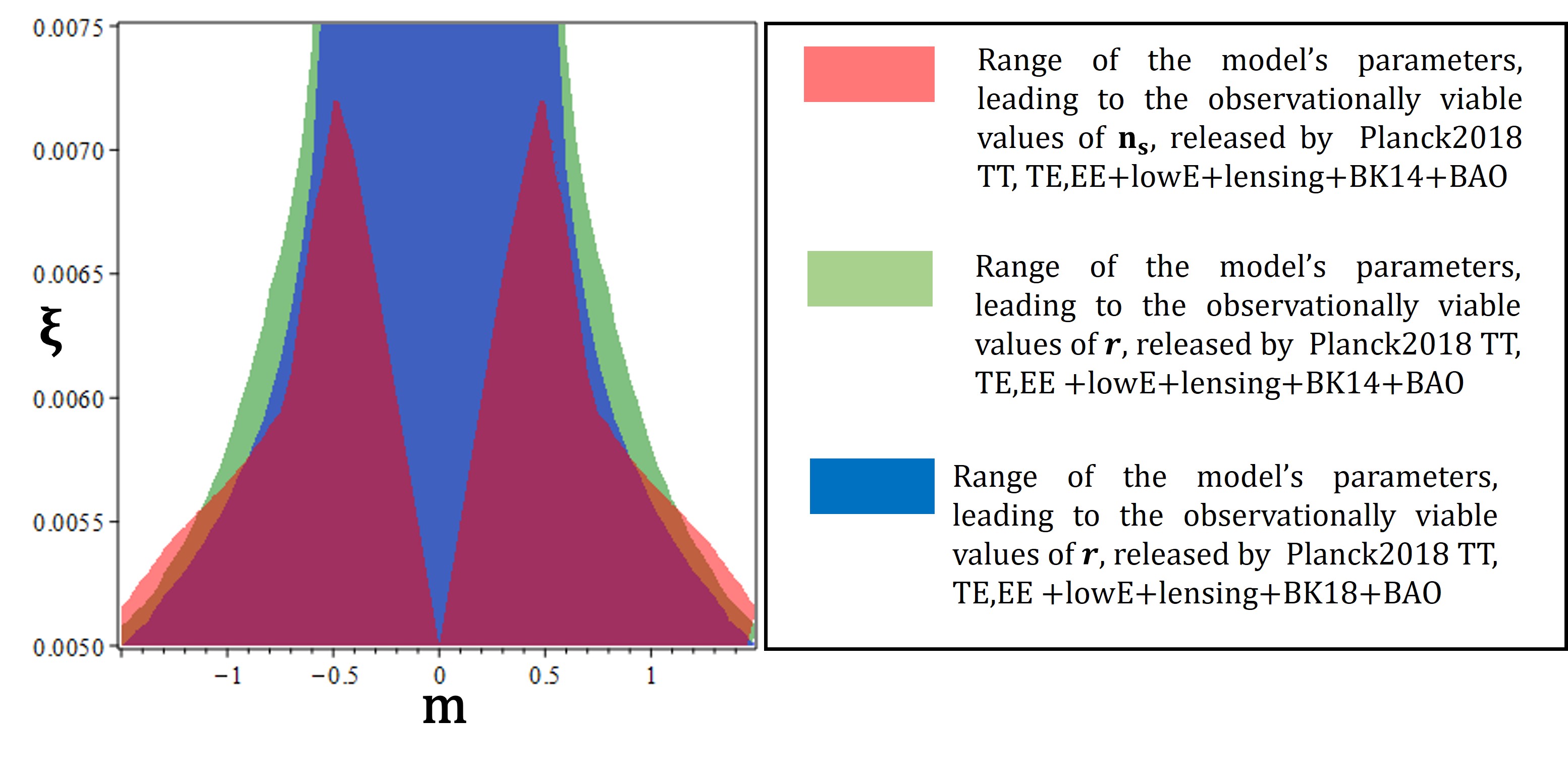}
	\caption{Ranges of the parameters $ \xi $ and $ m$ for $\lambda=A\phi^{n} $ model in the Jordan frame with constant $ n= 0.525$. In this case, the constraints on the NMC parameter is $ 0.0057 \leq \xi \leq 0.0072$, for $-1.5< m < 1.5$.}\label{figure.2}
\end{figure}

\begin{figure}[H]
	\centering
	\includegraphics[height= 7cm, width=14cm]{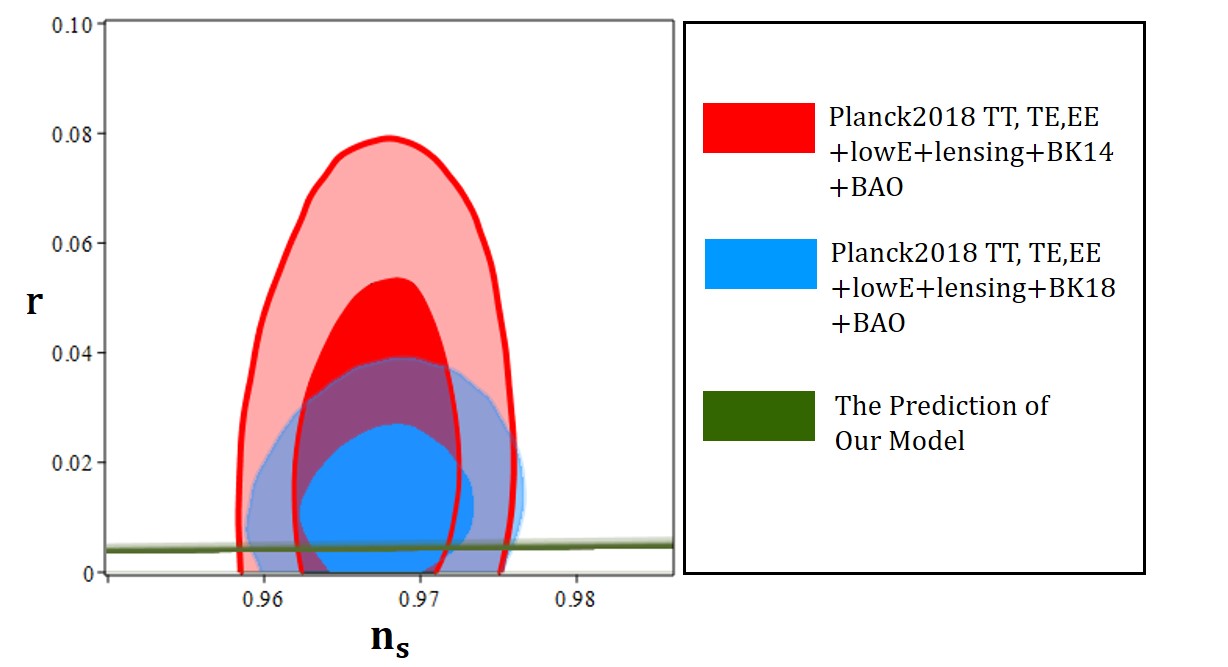}
	\caption{Tensor-to-scalar ratio $ r$ versus the scalar spectral index $ n_{s}$ plane for the Planck
		2018 and BICEP/Keck
		data for $\lambda=A\phi^{n} $ in the Jordan frame with constant $ m= \pm 0.505 $. 
	As shown in the figure, the ranges of the parameters $ \xi $ and $ n $ are consistent with Planck2018 and BICEP/Keck at $ 68\% $ and $ 95\% $ CL \cite{akrami2020planck, ade2021improved}. The green region representing our model's prediction agrees with the red and blue areas of the mentioned data set.}\label{figure.3}
\end{figure}

\begin{figure}[H]
	\centering
	\includegraphics[height= 7cm, width=14cm]{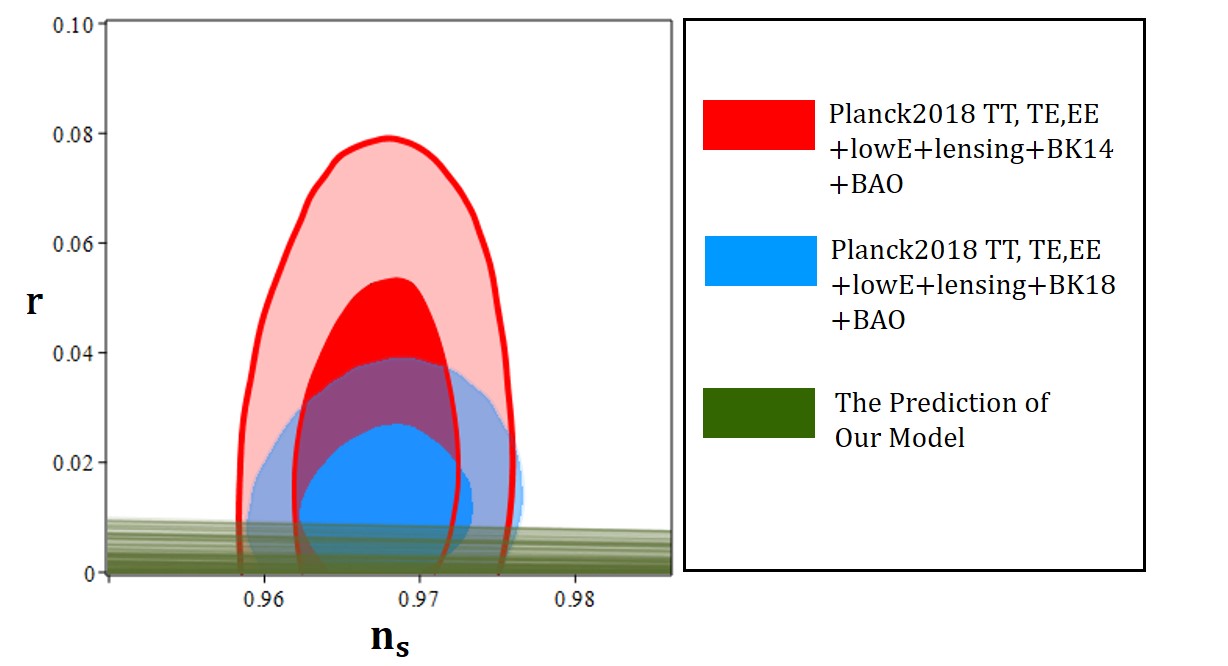}
	\caption{Tensor-to-scalar ratio $ r$ versus the scalar spectral index $ n_{s}$ for $\lambda=A\phi^{n} $ in the Jordan frame with constant $ n= 0.525$. As shown in the figure, the ranges of the parameters $ \xi $ and $ m $ are consistent with Planck2018 and BICEP/Keck at $ 68\% $ and $ 95\% $ CL \cite{akrami2020planck, ade2021improved}. The green region representing our model's prediction agrees with the red and blue areas of the mentioned data set.
	}\label{figure.4}
\end{figure}

\begin{figure}[H]
	\centering
	\includegraphics[height= 7cm, width=14cm]{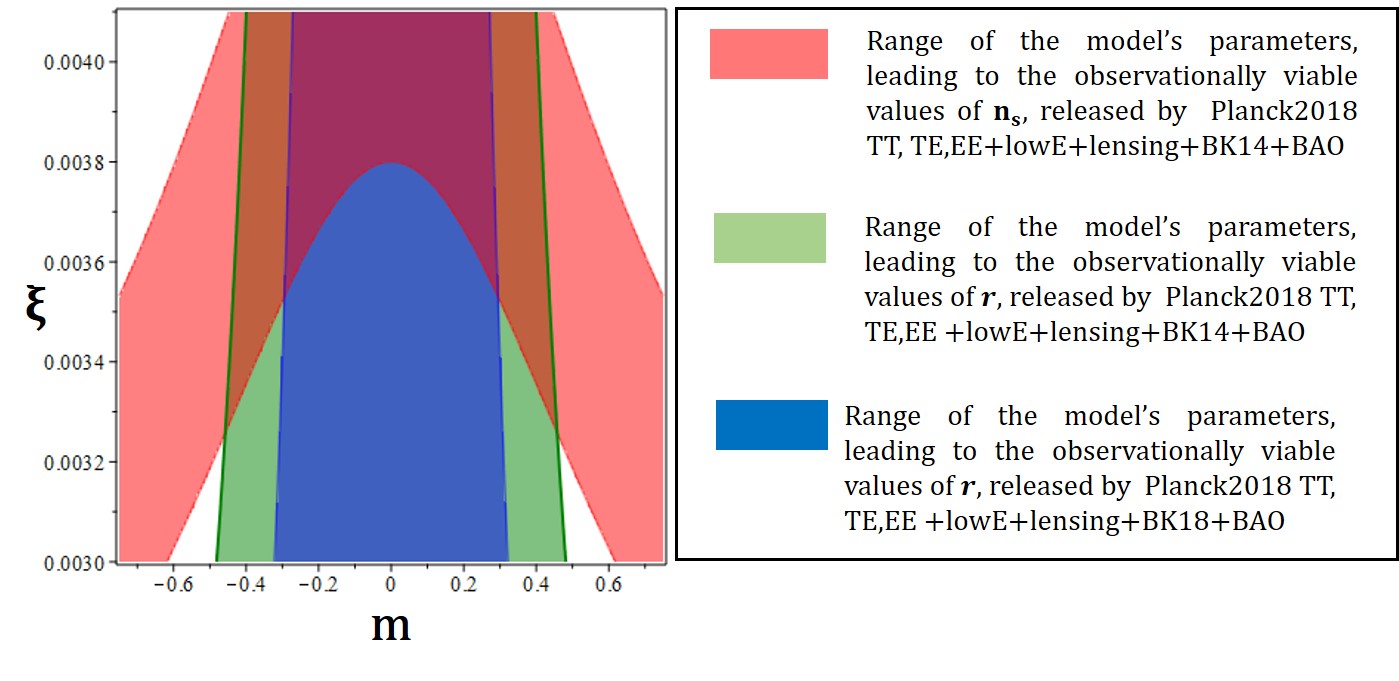}
	\caption{Ranges of the parameters $\xi $ and $m$ in the Einstein frame for the \emph{first order} of the slow-roll parameters. In this case for $ r<0.072$, the constraint on the NMC parameter $\xi$ is $ 0.0033 \leq \xi \leq 0.0047$, where $-0.4577<m<0.4577$.}\label{figure.5}
\end{figure}

\begin{figure}[H]
	\centering
	\includegraphics[height= 7cm, width=14cm]{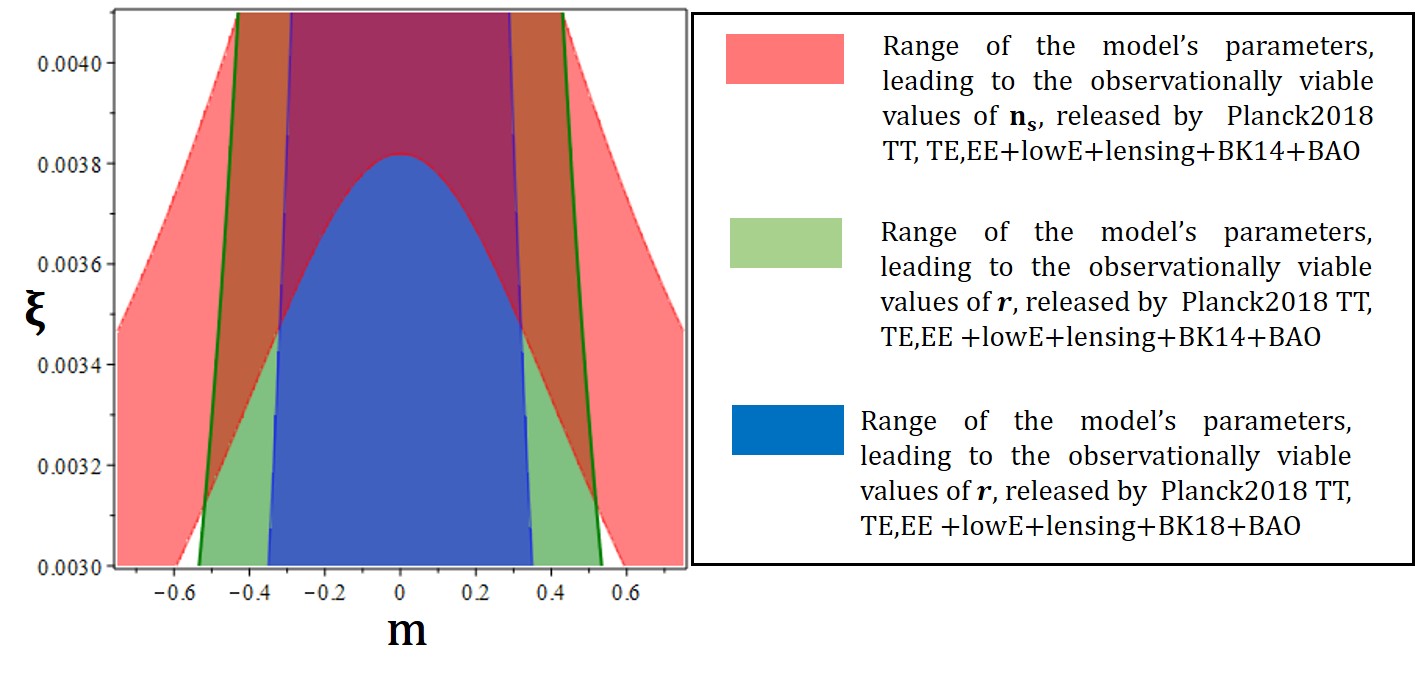}
	\caption{Ranges of the parameters $\xi $ and $m$ in the Einstein frame for the \emph{second order} of the slow-roll parameters. In this case for $ r<0.072$, the constraint on the NMC parameter is $0.0032 \leq \xi \leq 0.0047 $ for $-0.5183 <m< 0.5183$.}\label{figure.6}
\end{figure}

\begin{figure}[H]
	\centering
	\includegraphics[height= 7cm, width=14cm]{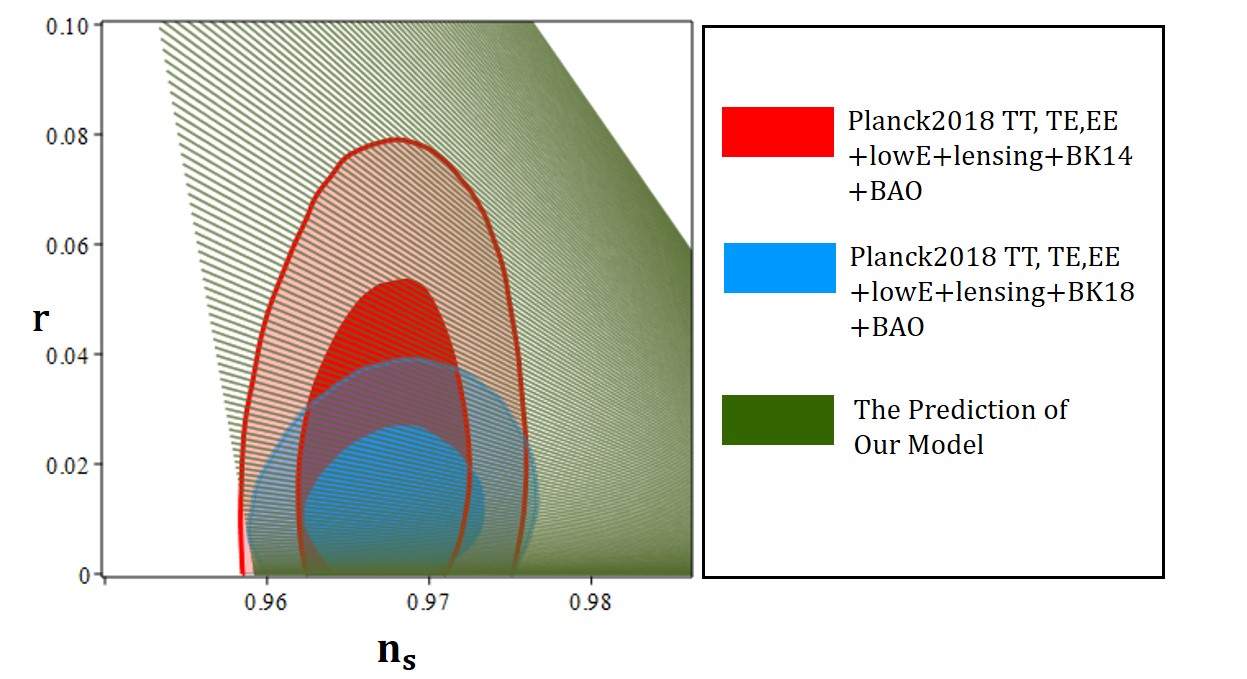}
	\caption{Tensor-to-scalar ratio $r$ versus the scalar spectral index $n_{s}$ in the Einstein frame for the \emph{first order} of the slow-roll parameters. As shown in the figure, the ranges of the parameters $ \xi $ and $ m $ are consistent with Planck2018 and BICEP/Keck at $ 68\% $ and $ 95\% $ CL \cite{akrami2020planck, ade2021improved}. The green region representing our model's prediction agrees with the red and blue areas of the mentioned data set.}\label{figure.7}
\end{figure}

\begin{figure}[H]
	\centering
	\includegraphics[height= 7cm, width=14cm]{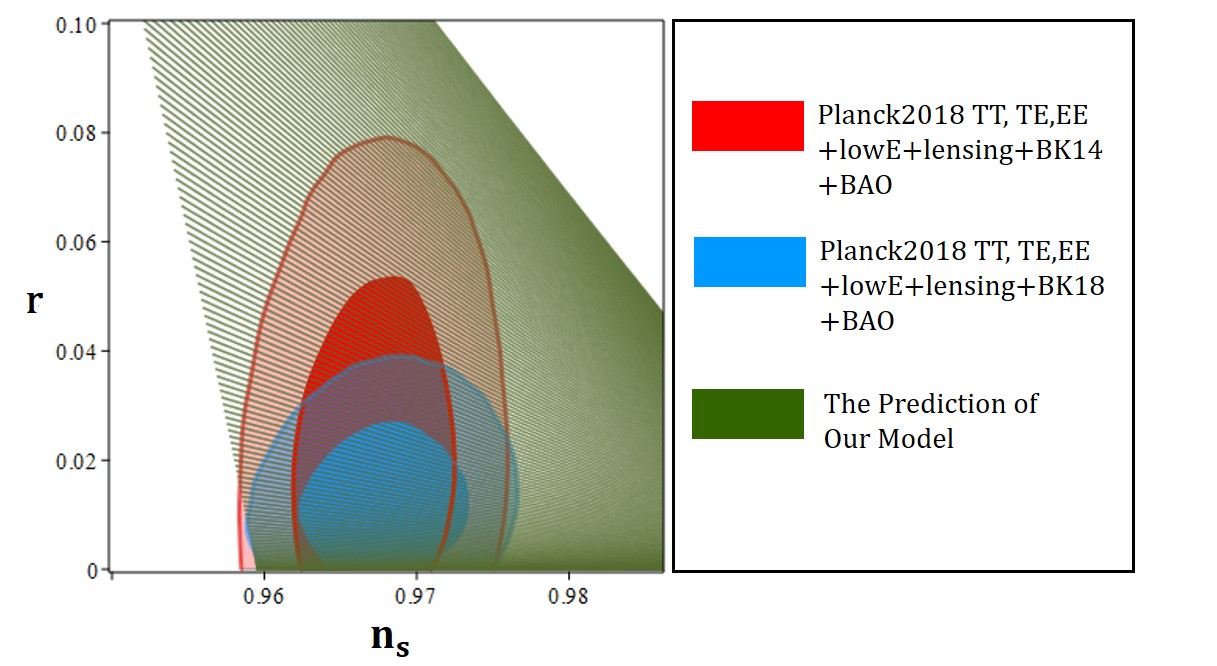}
	\caption{Tensor-to-scalar ratio $r$ versus the scalar spectral index $n_{s}$ in the Einstein frame for the \emph{second order} of the slow-roll parameters.
	As shown in the figure, the ranges of the parameters $ \xi $ and $ m $ are  consistent with Planck2018 and BICEP/Keck at $ 68\% $ and $ 95\% $ CL \cite{akrami2020planck, ade2021improved}. The green region representing our model's prediction agrees with the red and blue areas of the mentioned data set.}\label{figure.8}
\end{figure}

Using the Maple20 program, we get various sample values of the parameters $\xi$, $m$, and $n$ for $n_{s}=0.9658\pm 0.0038$, $r<0.072$, and $r<0.036$. Then, for each $r$, we choose one sample for comparison and to check for consistency with observational data. As we proceed, we face certain restrictions in both the Jordan and Einstein frames. Since the model has a wide parameter space, to constrain the NMC parameter, we use the observational constraints on $n_s$ and $r$, but since $m$ and $n$ are some essentially free parameters of the model, we fixed them on some appropriate values. As we explained, the appropriateness here means adaptation of those values of $m$ and $n$ that save the model in the confidence levels of the used data sets. In each of the Figures \eqref{figure.1} and \eqref{figure.2}, we show the results where the values of $m$ or $n$ are fixed, respectively, and a constraint is obtained on the parameter $\xi$ for the $\lambda=A\phi^{n}$ model in the Jordan frame. Figures \eqref{figure.3} and \eqref{figure.4} are given the $n_s-r$ for the same condition as Figures \eqref{figure.1} and \eqref{figure.2}. In Figure \eqref{figure.5}, ranges of the parameters $\xi$ and $m$ are given in the Einstein frame for the \emph{first order} of the slow-roll parameters. In the same way, in Figure \eqref{figure.6}, the ranges of the parameter space are plotted in the Einstein frame for the \emph{second order}. Similarly Figures \eqref{figure.7} and \eqref{figure.8} plot the $n_s-r$ for the first order and second order contributions. Table \eqref{table-1} displays the range of $n $ and $\xi$ parameters that we determine for two distinct cases: $r<0.072$ and $r<0.036$. The variable $ m $ has been held constant throughout the analysis. Table \eqref{table-2} provides an explicit ranges for the $ m $ and $ \xi $ parameters for $r<0.072$ and $r<0.036$, while keeping $ n $ constant. Tables \eqref{table-3} and \eqref{table-4} are constructed based on the ranges of parameter $m$ and $\xi$ for the first and second orders, for two distinct cases: $r<0.072$ and $r<0.036$. In summary:

\begin{itemize}
\item
In the Jordan frame and for $\lambda=A\phi^{n}$, we can see the results in Figures. We have ranges of parameters $\xi$, $ n $, and $ m $ for $r<0.072$ and $r<0.036$. For example, Figure \ref{figure.1} shows the range of parameters $\xi$ and $n$. If we choose $m=\pm 0.505$, then we find a constraint on the model's parameters as $0.0055\leq \xi \leq 0.0105$ and $n\geq0.525$. We also note that the constraints $r<0.072$ and $r<0.036$ are related to two different joint data sets: $r<0.072$ for Planck2018 \cite{akrami2020planck} and $r<0.036$ for BICEP/Keck \cite{ade2021improved}.

\item
Figure \ref{figure.2} shows the range of parameters $\xi$ and $m$ if we choose $n=0.525$ as a fixed parameter. So we have found a constraint on the model's parameters for $-1.5<m<1.5$ as $0.0057 \leq \xi \leq 0.0072$.

\item
Figures \ref{figure.3} and \ref{figure.4} show the tensor-to-scalar ratio $r$ versus the scalar spectral index $n_s$ for the $\lambda=A\phi^{n} $ model in the background of the Planck2018 and BICEP/Keck data at both $68\%$ and $95\%$ CL \cite{akrami2020planck, ade2021improved}.

\item
Based on these analyses, in Tables \ref{table-1} and \ref{table-2}, our model's parameters in the Jordan frame are summarized.
\end{itemize}

\begin{table}[H]
	\caption{Constraining model's parameters $n$ and $\xi $ (with fixed $m$) in confrontation with Planck2018 and BICEP/Keck collaborations joint data in the Jordan frame; $r<0.072$ for Planck2018 and $r<0.036$ for BICEP/Keck.}
	\label{table-1}
	\begin{center}
		\begin{tabular}{c c c c}
			\hline
			\hline
			$ r $ & $ m $ & $ n $& $\xi $ \\
		    	\toprule
			$ r <0.036 $ & $m= \pm 0.505$ & $ n\geq 0.418 $ & $ 0.0055 \leq \xi \leq 0.0105$ \\
			\hline
			$ r <0.072 $ & $m= \pm 0.505$ & $n\geq 0.525$ & $0.0054 \leq \xi \leq 0.0105$ \\
			\bottomrule
		\end{tabular}
	\end{center}
\end{table}

	\begin{table}[H]
		\caption{Constraining parameter $m$ with the fixed values of $n$ in confrontation with Planck2018 and BICEP/Keck collaborations joint data in the Jordan frame. The ranges of $\xi$ are chosen to be the same as the ranges in Table \eqref{table-1}.}
		\label{table-2}
		\begin{center}
			\begin{tabular}{c c c c}
			\hline
			\hline	
			$ r $&	$ n $ & $ m $ &$ \xi $ \\
			\toprule
		 $ r<0.036 $ & $n=0.525$ & $-1.5<m< 1.5$ & $0.0057\leq \xi \leq 0.0072$ \\
		 \hline
		 	$ r<0.072 $ & $n=0.525$ &  $-1.5<m< 1.5$  & $0.0055 \leq \xi \leq 0.0072$ \\
		 		\bottomrule
			\end{tabular}
		\end{center}
	\end{table}

Likewise, for the Einstein frame, we show figures and tables in two different orders of approximations and for two values of the tensor-to-scalar ratio $ r<0.036 $ and $ r<0.072 $.

\begin{itemize}
	\item
	In Figure \ref{figure.5}, we show the ranges of the parameters $ m $ and $ \xi$ in the Einstein frame for \emph{ the first order} of the slow-roll parameters. Our numerical analysis for the first order gives constraints on the model's parameters as $ -0.4577 <  m  < 0.4577 $ and $ 0.0033 \leq \xi \leq 0.0047 $ for $ r<0.072 $. Based on our analysis, we find constraint on the model's parameters as  $ -0.2944 < m < 0.2944 $ and $ 0.0035 \leq \xi \leq 0.0047$ for $ r<0.036 $.
	
	\item
	In the same manner, in Figure \ref{figure.6} for \emph{the second order} of the slow-roll parameters, we have found constraint's on the model's parameters as $ -0.5183 < m < 0.5183 $ and $ 0.0032 \leq \xi \leq 0.0047 $ for $ r<0.072 $ and also $ -0.3195 < m < 0.3195 $ and $ 0.0034 \leq \xi \leq 0.0047 $ for $ r<0.036$.

	\item
	  Figures \ref{figure.7} and \ref{figure.8} show the tensor-to-scalar ratio $ r$ versus the scalar spectral index $ n_{s}$ in the Einstein frame for the first and second orders of the slow-roll parameters, respectively. We find consistency with the Planck2018 \cite{akrami2020planck} and BICEP/Keck \cite{ade2021improved} data sets at $ 68\% $ and $ 95\% $ CL.

	 \item
	 In Tables \ref{table-3} and \ref{table-4}, we have summarized some constraints obtained in studying the Einstein frame for the first and second orders of the slow-roll parameters.
\end{itemize}

\begin{table}[H]
	\caption{Constraining model's parameters $m$ and $\xi $ in confrontation with Planck2018 and BICEP/Keck collaborations data in the Einstein frame in the first order.}
	\label{table-3}
	\begin{center}
		\begin{tabular}{c c c}
			\hline
			\hline
			$ r $	& $ m $ & $ \xi $ \\
			 \toprule
			$ r <0.036 $ & $ -0.2944 < m < 0.2944 $ & $ 0.0035 \leq \xi  \leq 0.0047 $ \\
			\hline
			$ r <0.072 $ & $ -0.4577 <  m  < 0.4577 $ & $ 0.0033 \leq \xi  \leq 0.0047 $ \\
			\bottomrule
		\end{tabular}
	\end{center}
\end{table}

\begin{table}[H]
	\caption{Constraining model's parameters $m$ and $\xi $ in confrontation with Planck2018 and BICEP/Keck collaborations data in the Einstein frame in the second order.}
	\label{table-4}
	\begin{center}
		\begin{tabular}{c c c}
			\hline
			\hline
			$ r $ & $ m $ & $\xi $ \\
			 \toprule
			$ r <0.036 $ & $ -0.3195 < m < 0.3195 $ & $ 0.0034 \leq \xi  \leq 0.0047 $ \\
			\hline
			$ r <0.072 $ & $ -0.5183 <  m  < 0.5183 $ & $ 0.0032 \leq \xi  \leq 0.0047 $ \\
			\bottomrule
		\end{tabular}
	\end{center}
\end{table}

As it is evident in our analysis, especially in the tables, although the differences between the first and second order results are small, but these differences are important since may have some traces on the issue of frames, that is, equivalency of two frames from physics point of view (mathematically these two frames are equivalent). In other words, taking into account the higher order contributions in slow-roll parameter in this setup, my shed light on the issue of frames definitely.

\section{Conclusion}\label{sec7}

As a model for explanation of early time cosmological inflation, a single scalar field inflationary model provides one of the best solutions both theoretically and phenomenologically. Also, for this model with suitable potentials, the most recent observations of the CMB are in excellent agreement with the predictions. In this paper, we have constructed a new version of the single field inflation where the scalar field is non-minimally coupled to the gravitational sector in a unimodular framework. Then, the cosmological aspects of this new theoretical extension are studied in detail. The main characteristic of the present model is that we consider the inflaton to be a generic, mathematical field without kinetic and potential terms in the action. Since the non-minimal coupling with the gravitational sector by itself produces an effective energy-momentum tensor, and also due to the unimodular nature of the model, this generic field has the capability to derive cosmological inflation. It is well-known that unimodular gravity lies in a subspace of general relativity. However, the cosmological constant is generally considered as an integration constant/Lagrange multiplier in this setup. In the present setup, a scalar field that is non-minimally coupled to gravity drives the cosmological inflation; that is, the field that is non-minimally coupled to unimodular gravity drives the cosmic inflation. Moreover, the value of the NMC parameter, $ \xi $, can be fixed in this setup in confrontation with recent observations. In our non-minimal unimodular scenario, we investigated the issue of cosmic inflation by analyzing the problem in both Jordan and Einstein frames in details. In the presence of the NMC, our analysis shows the possibility of natural exit from the inflation phase without additional mechanism. To conduct our investigation, firstly we derived the equations of motion in the Jordan frame in  non-minimal unimodular gravity and utilized the power law ansatz $\lambda= A \phi^{n}$ for Lagrange multiplier. Then we switched to the Einstein frame where the value of $\lambda$ is constant, in contrast to the Jordan frame. Then, we computed the spectra of inflationary perturbations like the scalar spectral index and tensor-to-scalar ratio. The parameters space of the model is wide and therefore to find some constraints on the important parameters such as $\xi$, we fixed some less important parameters to appropriate fixed values. Specially we constrained parameters $n$, $m$ and more importantly $\xi$ in this manner. The results are shown in Figures and Tables. Furthermore, we compared our results in two frames and in two orders of calculations in terms of the slow-roll parameters. The numerical analysis of the model parameters in two frames and then confronting the results with Planck and BICEP/Keck joint data shows that this non-minimal unimodular inflation is consistent with observations in some subspaces of the model parameter space. In this manner we obtained some novel and severe constraints on the model parameters, specially, on NMC parameter, $\xi$. A comparison between the results of calculations in the first and second orders of the slow-roll parameters and observing some tiny differences gives some traces of possibility to judge about perfect physical equivalency of these two mathematically and equivalent frames. To be more clarified, it seems that higher order calculations of the inflation observables in this non-minimal unimodular setup signals the possibility of non-equivalency of these two frames on physical ground. This issue is under study and will be reported in future. The power of the present model is that we consider a generic, mathematical scalar field without kinetic and potential terms in the Jordan frame action. The only source of the energy-momentum that drives the cosmological inflation is the effective energy-momentum coming from the NMC between the scalar field and the unimodular gravitational sector. As we have shown this model is successful in some respect, but has some issume in the some time. A more concrete analysis should consider the perturbation on the unimodularity factor $\lambda$, the issue of which we are going to report in the future. The other point in this respect is a complete analysis of non-minimal Higgs inflation in the background of unimodular gravity that is in progress and will be reported in the future.

\end{document}